\documentclass[final,5p,times,twocolumn,authoryear]{elsarticle}

\usepackage{amssymb}

\journal{International Journal of Heat and Mass Transfer}

\usepackage{graphicx}
\usepackage{subfigure}
\usepackage{colortbl}
\usepackage{textcomp}
\usepackage{caption}
\usepackage{amsmath,amsfonts}
\usepackage{array,floatflt}
\usepackage[active]{srcltx}
\usepackage{float}
\usepackage[flushmargin,hang]{footmisc}
\usepackage{url}
\usepackage{breqn}
\usepackage{natbib}
\usepackage{lipsum}

\newcommand\Pra{\mbox{$\mathcal{P}r$}}       
\newcommand\Nu{\mbox{$\mathcal{N}u$}}        
\newcommand\Ra{\mbox{$\mathcal{R}a$}}        

\newcommand{\dt}{\frac{\partial}{\partial t}}

\begin{document}

\begin{frontmatter}



\title{Dielectrophoretic force-driven convection in annular geometry under Earth's gravity}


\author[label1]{Torsten Seelig\corref{cor1}}
\ead{seelig@b-tu.de}
\author[label1]{Antoine Meyer}
\author[label2]{Philipp Gerstner}
\author[label1]{Martin Meier}
\author[label1]{Marcel Jongmanns}
\author[label2]{Martin Baumann}
\author[label2]{Vincent Heuveline}
\author[label1]{Christoph Egbers}

\address[label1]{Dept. of Aerodynamics and Fluid Mechanics, BTU Cottbus - Senftenberg, Germany}
\address[label2]{Engineering Mathematics and Computing Lab (EMCL),  Interdisciplinary Center for Scientific Computing (IWR), Heidelberg University, Germany}
\cortext[cor1]{Corresponding author}

\begin{abstract}
Context: A radial temperature difference together with an inhomogeneous radial electric field gradient is applied to a dielectric fluid confined in a vertical cylindrical annulus inducing thermal electro-hydrodynamic convection. \\
Aims: Identification of the stability of the flow and hence of the line of marginal stability separating stable laminar free (natural) convection from thermal electro-hydrodynamic convection, its flow structures, pattern formation and critical parameters. \\
Methods: Combination of different measurement techniques, namely the shadowgraph method and particle image velocimetry, as well as numerical simulation are used to qualify/quantify the flow. \\
Results: We identify the transition from stable laminar free convection to thermal electro-hydrodynamic convective flow in a wide range of Rayleigh number and electric potential.
The line of marginal stability found confirms results from linear stability analysis.
The flow after first transition forms a structure of axially aligned stationary columnar modes.
We experimentally confirm critical parameters resulting from linear stability analysis and we show numerically an enhancement of heat transfer.
\end{abstract}


\end{frontmatter}


\section{Introduction}\label{Introducion}
Planetary flows are subjected to a conservative central force field and heat transfer due to several mechanisms like radiation, thermic conduction or convection.
Containment in engineering often provides thermal insulation.
An encapsulated set-up is the cask for storage and transport of radioactive material 'castor'.
Another example of such enclosures is a heat exchanger system.
There, an improvement of heat transfer via efficient enhancement is of general interest due to its benefits by low operational costs and due to sustainable use of energy.
In order to model those flows in the laboratory, we want to focus here on the method of applying a radially inhomogeneous electric field superposed on a radial temperature difference, known as thermal electro-hydrodynamic 'TEHD' driven convection~\citep{Bergles:1998,Marucho:2013,Yoshikawa:2013,Futterer_Dahley:2016}.

When an electric field $\mathbf{E}=-\nabla \Phi$, with electric potential $\Phi$, is applied a dielectric fluid is subjected to the electric body force~\citep[see, for example,][]{Stratten:1941,Landau_Lifshitz:1960,Pohl:1978,Landau_Lifshitz:1984}
\begin{eqnarray}\label{eq:electric_force}
\mathbf{F}_{E} & = & \mathbf{F}_{C}+\mathbf{F}_{DEP}+\mathbf{F}_{ES} \\ \notag
 & = & \rho_{E}\mathbf{E}-\frac{1}{2}\mathbf{E}^{2}\nabla\epsilon+\frac{1}{2}\nabla\left[\rho\left(\frac{\partial\epsilon}{\partial\rho}\right)_{T}\mathbf{E}^{2}\right], 
\end{eqnarray}
where $\rho_{E}$ is the free charge density, $\epsilon=\epsilon_{0}\epsilon_{f}$ is the dielectric constant, the product of the permittivity of free space $\epsilon_{0}$ and fluid's relative permittivity $\epsilon_{f}$, and $\rho$ is the density of the fluid.
The first term is identified as Coulombic, the second as dielectrophoretic and the third as the electrostrictive force.
In general, the gradient force $\mathbf{F}_{ES}$ has no contribution to the flow field and will be considered in the momentum equation as an additional term in the pressure gradient force~\citep{Yoshikawa:2013,Mutabazi:2016,Zaussinger:2018}.
Applying a direct current (d.c.) electric field exerts the Coulomb force.
On the other hand, applying an inhomogeneous alternating current (a.c.) electric field with a frequency much higher than the inverse of the charge relaxation time of the dipole molecules prevents from free charges accumulation.
Dielectric fluids have relaxation times of about 10-100s.
Therefore, in an a.c. electric field with 60 Hz or higher frequency the Coulomb force vanishes and the dielectrophoretic force $\mathbf{F}_{DEP}$ is dominant.
Thus the dielectrophoretic (DEP) force has been remained in focus up to now, e.g. very recently in~\citet{Laohalertdecha:2007,Marucho:2013,dissDahley:2014,Travnikov:2015,Travnikov:2016,Futterer_Dahley:2016,Meyer:2017a,Meyer:2018,Meier:2018,Zaussinger:2018}.
The dielectrophoretic force depends on $\mathbf{E}^{2}$ instead of $\mathbf{E}$ and is therefore independent of the direction of $\mathbf{E}$.
If the frequency of the electric potential $\Phi$ is high enough it can be time-averaged over a period of the electric field.
Then the imposed electric potential $\sqrt{2}V_{0}\sin(2\pi ft)$ can be replaced by its effective value $V_{0}$.
\citet{Turnbull_Melcher:1969} found this assumption predicts successfully the onset of the thermal electro-hydrodynamic convection.

The problem of onset of convective instability of a dielectric fluid confined in a concentric vertical annulus subjected to Archimedean buoyancy force due to a radial temperature difference and Earth's gravity, and to dielectrophoretic force due to a radial alternating electric field has been studied by many authors experimentally~\citep{Smylie:1966,Chandra_Smylie:1972,dissDahley:2014,Futterer_Dahley:2016,Meyer:2017a}, theoretically~\citep{Takashima:1980,Takashima:1984,Stiles_Kagan:1993,Meyer:2017b} and numerically~\citep{Takashima:1984,Smieszek:2008}.
Consider a vertical cylindrical annulus with an inner heated cylinder maintained at temperature $T_{1}$ and an outer cooled cylinder maintained at temperature $T_{2}$.
The vertical annulus is of height $h$ with adiabatic top and bottom boundaries.
Without an electric field gradient, the temperature difference $\Delta T = T_{1} - T_{2}$ induces laminar free (natural) convection in the gap.
Superposing an a.c. electric field to the dielectric fluid presenting an electric permittivity gradient~\citep{Stratten:1941,Pohl:1978,Landau_Lifshitz:1984} the dielectrophoretic force is induced.
The inhomogeneity of the a.c. electric field gradient is provided by the curvature of the cylindrical annulus. 
Consequently there exists a radial directed central force field acting on the dielectric fluid which can be seen as resulting from the effect of an electric gravity~\citep[see, for example,][]{Chandra_Smylie:1972,Travnikov:2015,Mutabazi:2016,Futterer_Dahley:2016,Meyer:2017b}.
Comparable to Archimedean buoyancy acting in axial direction it leads to 'dielectrophoretic' buoyancy acting in the radial direction.
Dependent on the strength of the electric gravity and hence on electric potential there is a competition between Archimedean and dielectrophoretic buoyancy.
\citet{Yoshikawa:2013} investigated DEP force-driven convection in annular geometry and showed the basic electric gravity is centripetal for a warm inner cylinder and a cold outer cylinder.  
They found that, except for cylindrical annulus with low curvature, where perturbation of electric gravity plays a significant stabilising role, the critical parameter remains in the vicinity of its value for the classical Rayleigh B\'{e}rnard problem (\Ra$_{C}=1708$). 
This suggests the instability is driven by the same mechanism as in the gravity-driven ordinary thermal convection.
\citet{Meyer:2017a} reported the flow after the first transition is axially aligned and consists of stationary columnar modes.
\citet{Meier:2018} demonstrated experimentally the existence of those columnar modes for a few tuple of ($\Delta T,V_{0}$).

The present study focuses on the identification of the line of marginal stability where the flow undergoes a transition from stable laminar free (natural) convective flow to thermal electro-hydrodynamic convection.
Furthermore, we want to identify flow patterns and their spatial and temporal properties.
In order to achieve the objectives, we use different experiment cells described in section~\ref{experimental-set-up}.
Combination of two different measurement techniques, namely the shadowgraph method and particle image velocimetry (PIV), gains a better understanding of flow pattern and its stability.
Both are described in detail in section~\ref{Measurement-techniques}.
When PIV is applied to a dielectric fluid subjected to the DEP force the question arises if the used tracer particles are able to follow the flow field.
The influence of the DEP force on particle movement is considered in detail in section~\ref{slip}.
The numerical model used to obtain three-dimensional flow properties is introduced in section~\ref{numerical-model} followed by section~\ref{protocol} explaining how-to collect data and how we do post-processing.
Section~\ref{results} is devoted to results.
We show experimentally a regime diagram spanned by the non-dimensional Rayleigh number and electric potential.
We experimentally identify the line of marginal stability for the experiment cells of different aspect ratio and confirm linear stability analysis from~\citet{Meyer:2017a}.
The flow structure in the stable laminar free convective and in the thermal electro-hydrodynamic convective regime will be described in section~\ref{regimes}.
A discussion of pattern formation and growth follows in section~\ref{growth}.
Furthermore, we derive a regime diagram based on azimuthal wavenumber spanned by non-dimensional Rayleigh number and electric potential.
We also confirm critical parameters like vertical, azimuthal wavenumber and critical frequency found by linear stability analysis~\mbox{\citep{Meyer:2017a}}.
The enhancement of heat transfer is described in section~\ref{sec:heat_transfer}.

\section{Experimental set-up}\label{experimental-set-up}
\begin{figure*}
\centering
\includegraphics[width=\textwidth]{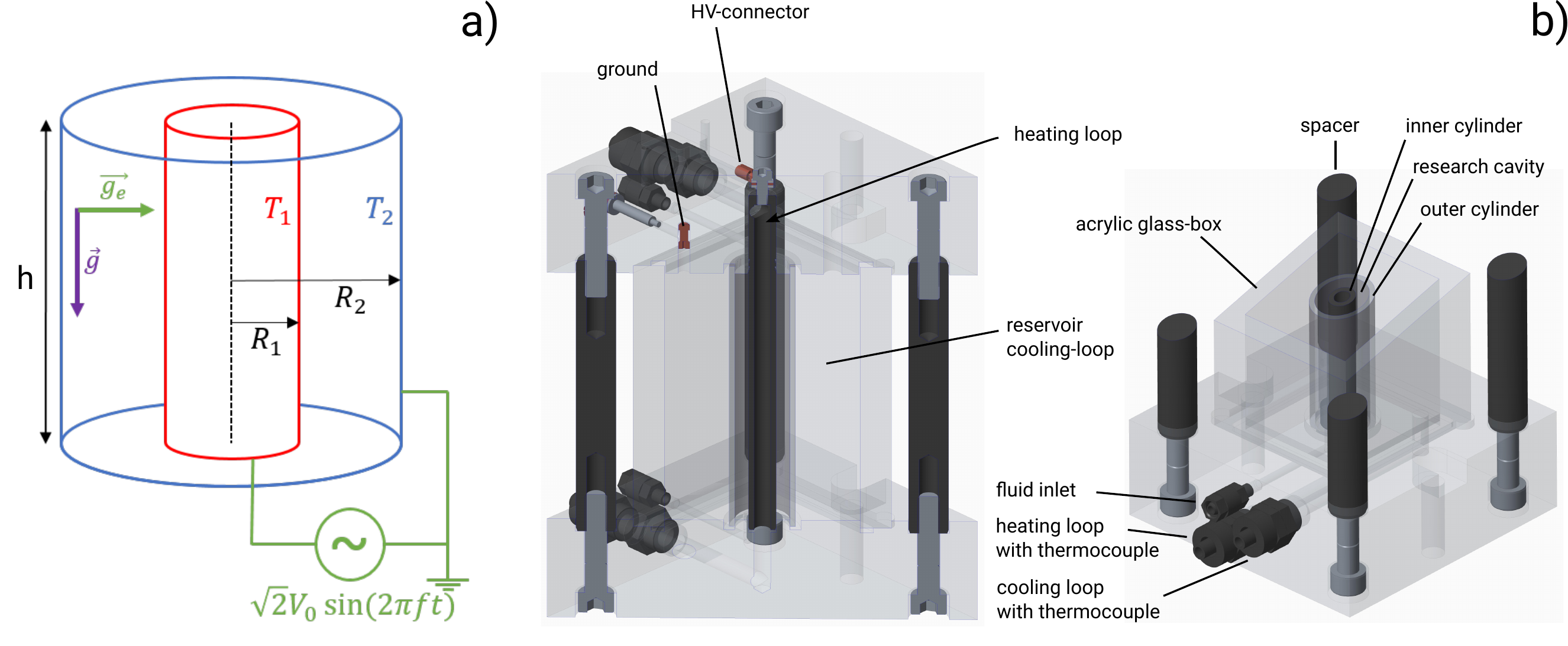}
\caption{(a) Schematic representation of the geometry of the experiment cell and the direction of applied forces. The inner cylinder is heated and electric potential is applied to it. The outer cylinder is cooled and connected to ground. (b) 3D mechanical drawing of the experimental set-up.}
\label{fig:ExpSetup}
\end{figure*}

This section briefly describes the laboratory experiment.
A detailed description is written in~\citet{Meyer:2017a} and~\citet{Meier:2018}.
The experiment consists of two concentric axially aligned cylinders (see figure~\ref{fig:ExpSetup}).
The gap in-between is filled with silicone oil Elbesil B5, physical properties see table~\ref{table:propOil}.
The inner cylinder has radius $R_{1}=5$ mm, the outer cylinder has radius $R_{2}=10$ mm and both have a height of $h=100$ or 300 mm, resulting in a gap width of $d=R_{2}-R_{1}=5$ mm, a radius ratio of $\eta=R_{1}/R_{2}=0.5$, an aspect ratios of $\Gamma=h/d=20$ or 60 and a curvature of $\beta=(R_{2}-R_{1})/R_{1}=1/\eta-1=1$.
For the use of the shadowgraph method inner and outer cylinder is made out of aluminium and both are anodised to suppress parasitic reflections.
Particle image velocimetry requires a transparent outer cylinder made out of borosilicate glass (wall thickness 2 mm).
The glass surface provides a wafer-thin semiconductor layer with resistance $R=10^{3}-10^{6}\,\Omega$.
The cylinder is cross-edge coated on both sides with 'transparent conductive oxide' (TCO).
TCO is electrically conductive, invisible, antistatic and usually consists of indium tin oxide (ITO), antimony tin oxide (ATO) or aluminium zinc oxide (AZO).
An alternating high-voltage electric potential $\sqrt{2}V_{0}\sin(2\pi ft)$ with $f=200$ Hz and $0.5\leq V_{0}\leq7$ kV is applied to the inner cylinder whilst the outer cylinder is connected to the ground.  
The annulus is placed and centred in a transparent acrylic glass-box and closed by a lower/upper lid also made of acrylic glass.
Cooling of the outer cylinder is realised by pumping silicone oil Elbesil B5 into the glass-box via in- and outlets, additionally it avoids optical distortions due to different refraction indices and curvature of the cylinders.
Heating of the inner cylinder is realised by pumping silicone oil Elbesil B5 via in/outlets into the inner cylinder.
For experiments with the shadowgraph method applied we heat and cool with a custom-built technique very detailed described in~\citet{Meyer:2017a} and~\citet{Meier:2018}.
For experiments using particle image velocimetry, we heat and cool with two thermostats.
We control and measure the applied temperature difference $\Delta T=T_{1}-T_{2}$ with thermocouples integrated into the inlet of the loops.
All fluid loops and in/outlets have been sealed against thermal radiation.

\begin{table*}[t]
\centering
\begin{tabular}{rlcc}
\hline\noalign{\smallskip}
\multicolumn{2}{c}{property} & unit & Elbesil B5  \\[3pt]
\hline\noalign{\smallskip}
density & $\rho_{0}$ & kg/m$^{3}$ & 923$\pm1.5$ \\
kinematic viscosity & $\nu$ & m$^{2}$/s & $5.54\pm0.06\cdot10^{-6}$ \\
thermal diffusivity & $\kappa$ & m$^{2}$/s & $7.74\cdot10^{-8}$ \\
thermal expansion coefficient & $\alpha$ $(0\hdots 150^{\circ}\rm{C})$ & 1/K &  $1.08\cdot10^{-3}$ \\
thermal conductivity & $\lambda$ & W/(mK) & 0.133 \\
specific heat capacity & $c_{p}$ & J/(kgK) & 1630 \\ 
electrical conductivity & $\sigma$ & ($\Omega$m)$^{-1}$ & $>10^{-12}$ \\
rel. permittivity & $\epsilon_{f}$ (100 Hz) & $-$ & 2.7 \\
dielectric variability & $\gamma$ & 1/K & $1.065\cdot10^{-3}$ \\
permittivity of free space & $\epsilon_{0}$ & A$^{2}$s$^{2}$/(Nm$^{2}$) & $8.8543\cdot10^{-12}$ \\
Earth's gravity & $g$ & m/s$^{2}$ & 9.80665 \\
\noalign{\smallskip}\hline
\end{tabular}
\caption[Physical properties.]{Physical properties of silicon oil Elbesil B5 at $25^{\circ}\rm{C}$ and some physical constants.
}
\label{table:propOil}
\end{table*}

\section{Measurement techniques}\label{Measurement-techniques}
\subsection{Shadowgraph method}\label{Shadowgraph}
Visualisation of the flow of cell with aspect ratio $\Gamma=20$ is realised with the shadowgraph method \citep{Merzkirch:1987,Schoepf:1996,Meyer:2017a,Meyer:2017b,Meier:2018}.
Shadowgraph is a very simple technique and provides flow visualisation of very high quality.
It is a non-disturbing method using the fact that the refractive index of a fluid depends on the density and hence on temperature.
An applied temperature difference to the fluid implies an inhomogeneous variation of the refractive index inside the fluid.
Monochromatic red (635 nm) telecentric light originating from a LED panel illuminates the annulus from below.
The intensity of the transmitted light is altered with respect to the incoming light, it is refracted towards colder regions resulting in a new pattern of light intensity at the top level of the annulus.
A mirror mounted on top of the experiment cell redirects the light.
A camera focused on top level of the annulus captures the density changes.
The $\mu$Eye camera (IDS UI-5550SE) has a CMOS sensor with a resolution of $1600\times 1200$ px, a rolling shutter and a dynamic range of 12 bit.
Additionally, we mounted an Edmund Optics 59-872 lens with focal length $f=35$ mm.
We record snapshots as well as time-series' with a framerate of \mbox{10 fps}.
Only the red channel of the RGB images is used for post-processing.
To enhance the contrast (i) the measurement is normalised with a reference image captured with temperature difference applied which leads to the state of ’natural convection’ and (ii) we use a false colour representation.
Since refractions are integrated over the height of the cell and the method requires global variations in the refractive index the shadowgraph method is more qualitative.
Furthermore, telecentric rays emitted below the annulus are refracted dependent on fluid properties and applied temperature difference more or less towards colder regions.
A region without light intensity around the inner cylinder appears, whose radial extent depends on the vertical extent of the annulus.
Therefore, the shadowgraph method is only applicable to the experiment cell with aspect ratio $\Gamma=20$.

\subsection{Particle image velocimetry (PIV)}\label{PIV}

A second measurement technique we use is the well known PIV technique to measure flow velocities~\citep{Tropea:2007,Adrian_and_Westerweel:2010,Westerweel:2013}.
We mix Potters hollow glass sphere (HGS) particles made of borosilicate glass with density $\rho_{p}=1100$ kg/m$^{3}$, averaged radius $R_{p}=5\,\mu$m and relative permittivity $\epsilon_{p}=4.6$ into the fluid.
Furthermore, the setup consists of a continuous monochromatic green (532 nm) diode-laser.
The 50 mW laser module has a lens (divergence angle 110$^{\circ}$) with fix and uniform focus producing a very homogeneous laser line that fades out towards the ends (Gaussian light distribution), resulting in a so-called 'powerline' of about 90$^{\circ}$and thickness of 1.4 mm.
We adjusted the laser module carefully, with the result that the laser illuminates particles in an (r,z)-plane.
The GigE camera (Imaging Source DMK 33GX174) has a CMOS pregius sensor with a resolution of $1920\times 1200$ px (2.3 MP) up to 50 fps, a global shutter and a dynamic range of 12 bit.
Additionally, we mounted an Imaging Source 5MP lens with focal length $f=12$ mm and iris range 1:1.4.
The camera is placed and centered at mid-height normal to the laser plane.
We kept the distance between lens and laser plane as short as possible to optimise the vertical resolution which depends on the used experiment cell.
The motion of illuminated particles is recorded as 15-minute time-series with a framerate of \mbox{10 fps} and stored in compressed movie files with the help of the software package IC Capture 2.4.
There is an advantage within the software to adjust/optimise the field of view in order to reduce the movie file size.
Therefore, recorded images have resolution $200\times1920$ px in case of the cell with aspect ratio $\Gamma=20$ and $100\times 1920$ px in case of the cell with aspect ratio $\Gamma = 60$.

We calibrated the field of view with a chessboard pattern placed and adjusted to the laser-light sheet.
From each record, we extract grey scale images using a \textit{Unix} command line tool to convert multimedia files between formats called 'ffmpeg'.
Distortions appearing due to refraction, perspective view and camera's lens distortion we correct by applying a 'Polynomial' distortion method.
The polynomial distortion maps pairs of source control points ($X_{S},Y_{S}$) basically the checkerboard pattern nodes to destination control points ($X_{D},Y_{D}$) using a standard polynomial equation of order five. Destination grid points have been chosen to achieve a radial/vertical resolution of 20/17 px/mm. The method uses all the control point pairs given to calculate the appropriate coefficients ($C_{iX},C_{iY}$), being $i = 0,\hdots,21$,
\begin{eqnarray}
X_{D} & = & C_{21X}X^{5}_{S} + C_{20X}Y^{5}_{S} + \hdots + C_{3X}X_{S}Y_{S}  + C_{2X}X_{S}  \notag \\
      &  & + C_{1X}Y_{S} + C_{0X},  \notag \\
Y_{D} & = & C_{21Y}X^{5}_{S} + C_{20Y}Y^{5}_{S} + \hdots + C_{3Y}X_{S}Y_{S}  + C_{2Y}X_{S}  \notag \\
      &  & + C_{1Y}Y_{S} + C_{0Y}.
\end{eqnarray}

For efficiency we use the library 'convert' of 'ImageMagick's' software package.
The velocity is analysed in three iteration steps using the MatPIV 1.6 toolbox~\citep{Sveen:2004} with an interrogation window size of \mbox{$32\times128$ px}, \mbox{$16\times64$ px} and \mbox{$8\times32$ px}, respectively.
The interrogation windows have an overlap of 50\%.
It results in a radial resolution of approximately 0.2 mm for both cells and a vertical resolution of approximately 0.9 or 2.6 mm in case of the experiment cell with aspect ratio $\Gamma=20$ or 60.

\subsubsection{Ability of tracer particles to follow the flow field}\label{slip}
\begin{table*}[t]
\centering
\begin{tabular}{ccccc}
\hline\noalign{\smallskip}
temperature $\vartheta$ ($^{\circ}\rm{C}$) & density $\rho_{f}$ (kg/m$^{3}$) & kinematic viskosity $\nu$ (m$^{2}$/s) & \multicolumn{2}{c}{sedimentation velocity (mm/s)}\\
 & & & $R_{p} = 5\;\mu$m & $R_{p} = 10\;\mu$m\\[3pt]
\hline\noalign{\smallskip}
21.66	& 925.8$\pm$1.5	& 5.9484$\pm$0.0678$\cdot 10^{-6}$ & 1.7$\cdot 10^{-3}$ & 6.9$\cdot 10^{-3}$\\
25.78	& 923.0$\pm$1.5	& 5.5363$\pm$0.0631$\cdot 10^{-6}$ & 1.9$\cdot 10^{-3}$ & 7.5$\cdot 10^{-3}$\\
28.30	& 920.9$\pm$1.5	& 5.3001$\pm$0.0604$\cdot 10^{-6}$ & 2.0$\cdot 10^{-3}$ & 8.0$\cdot 10^{-3}$\\
\noalign{\smallskip}\hline
\end{tabular}
\caption[Temperature dependence.]{Density and kinematic viscosity of silicon oil Elbesil B5 as function of temperature.}
\label{table:tempdep}
\end{table*}

Without electric potential, the weight force or gravitational force of particles suspended in a fluid is balanced by buoyancy force, due to particles density relative to the fluid and frictional force due to viscosity.
If one considers spherical particles of radius $R_{p}$ the Stokes' theorem describes the frictional force.
Rearranging the equation describing the equilibrium of forces results in the density contrast sedimentation velocity
\begin{eqnarray}
 \mathbf{v_{p}} & = & -\frac{2}{9}\frac{gR_{p}^{2}}{\nu}\left( \frac{\rho_{p}}{\rho_{f}} - 1 \right)\mathbf{e_{z}},
\end{eqnarray}

where $g$ is Earth's gravity, $\rho$ is the density whereas the index $p$ denotes a particle and the \mbox{index $f$} the surrounding fluid and $\nu$ is the kinematic viscosity.
Because $\rho_{p}>\rho_{f}$ is valid, the force leads to a relative movement of the particle in direction of Earth's gravity.
For a given temperature we measured the density and the kinematic viscosity with a viscosimeter (Anton Paar SVM3000) and summarised some particular values, valid for a typical temperature difference $\Delta T=7$ K, in table~\ref{table:tempdep}.
With the used particles with density $\rho_{p}=1100$ kg/m$^{3}$ we calculate the sedimentation velocity for two different particle radii, the averaged size $R_{p}=5\,\mu$m and maximum size $R_{p}=10\,\mu$m.
Values are also shown in table~\ref{table:tempdep}.
The values let us conclude that sedimentation velocity due to the density contrast of a particle relative to the fluid is negligible in a wide range of temperature and they are compatible with silicone oil.

Particles suspended in a dielectric fluid 'feel' an additional force, mainly due to particle polarisation relative to the fluid.
With a few assumptions, this effect can be expressed with the permittivity difference of particle material and fluid as shown in the following.
The dipole part of the dielectrophoretic force $\mathbf{F_{DEP}}$, acting on a particle with volume $V_{p}$, can be written as
\begin{eqnarray}\label{eq:Fdep}
  \mathbf{F_{DEP}} & = & -2\pi\epsilon_{0}\epsilon_{f}R_{p}^{3}\Re(f_{CM})\mathbf{\nabla \mid E\mid}^{2},
\end{eqnarray}

where $\epsilon_{0}$ is the permittivity of free space, $\epsilon_{f}$ the relative permittivity of the medium, $R_{p}$  the particle radius, $\Re(f_{CM})$ the real part of the relative particle polarization given by the frequency-dependent Clausius-Mosotti factor and $\mathbf{\nabla \mid E\mid}^{2}$ the gradient of the (squared) electrical field. 
For the derivation of eq.~\eqref{eq:Fdep} see~\citet{Pohl:1978} or~\citet{Jones:1995}.
The Clausius-Mosotti factor is defined as
\begin{eqnarray}\label{eq:CM}
 f_{CM} & = & \frac{\epsilon_{p}^{*}-\epsilon_{f}^{*}}{\epsilon_{p}^{*}+2\epsilon_{f}^{*}},
\end{eqnarray}

With $\epsilon^{*}$ the complex permittivity $\epsilon^{*}=\epsilon-{\mathrm i}\sigma\omega^{-1}$ of the particular material is expressed.
There ${\mathrm i}^{2}=-1$ is the imaginary unit, $\sigma$ the electrical conductivity and $\omega=2\pi f$ the field frequency. 
Ideally, the electric field should have frequencies $f$ high enough to suppress the influence of the Coulomb force~\citep{Travnikov:2004} and so that of the electrical conductivity.
Thus the dielectrophoretic force $\mathbf{F_{DEP}}$ caused by an alternating high voltage field acting on a spherical particle with radius $R_{p}$ and permittivity $\epsilon_{p}$ suspended in a medium with permittivity $\epsilon_{f}$ is given by
\begin{eqnarray}\label{eq:Fdep_fin}
 \mathbf{F_{DEP}} & = & 2\pi R_{p}^{3}\epsilon_{0}\epsilon_{f}\frac{\epsilon_{p}-\epsilon_{f}}{\epsilon_{p}+2\epsilon_{f}}\mathbf{\nabla \mid E\mid}^{2}.
\end{eqnarray}

The sign of the force depends only on the sign of the term $\epsilon_{p}-\epsilon_{f}$, the difference of the permittivities.
If $\epsilon_{p}>\epsilon_{f}$ is valid, the force leads to a relative movement of the particle in direction of the field gradient.
If the permittivity of the particle is less than the permittivity of the surrounding medium, the relative movement takes place in the opposite direction, contrary to the field gradient.
The particular formulation of $\mathbf{F_{DEP}}$ for the cylindrical geometry can be found in the subsequent sections.

As the eq.~\eqref{eq:Fdep_fin} tells, the optimal case would be the usage of particles with a permittivity equal to that of the surrounding medium, because no relative movement due to the dielectrophoretic force would occur and particles could be regarded as frozen into the fluid.
Because this is most likely not reachable in practice, the difference should be as small as possible.
On a first glance also a reduction of the particle radius $R_{p}$ seems to be helpful to reduce the force.
However, if the size is reduced, also the mass of the particle will be reduced and the force per particle (the acceleration) does not change.
Both expressions depend on the same power of $R_{p}$.
Hence the permittivity is the only free parameter for optimizations.

All experiments use an (inhomogeneous) alternating electrical field that can be described with the following expression
\begin{eqnarray}\label{eq:eP}
 \mathbf{E} & = & -\frac{\sqrt{2}V_{0}\sin(2\pi ft)}{r\ln(\eta)}\mathbf{e_{r}},
\end{eqnarray}
where $\sqrt{2}V_{0}$ is the amplitude of the alternating high voltage with its effective value $V_{0}$.

Combining eq.~\eqref{eq:Fdep_fin} for the dielectrophoretic force and eq.~\eqref{eq:eP} for the electrical field, the expression for the dielectrophoretic force in cylindrical geometry can be concretised and results in
\begin{eqnarray}\label{eq:Fdep_fin_cyl}
 \mathbf{F_{DEP}} & = & -2\pi \frac{R_{p}^{3}}{r^{3}}\epsilon_{0}\epsilon_{f}\frac{\epsilon_{p}-\epsilon_{f}}{\epsilon_{p}+2\epsilon_{f}}\left( \frac{\sqrt{2}V_{0}}{\ln(\eta)} \right)^{2}\sin^{2}(2\pi ft)\mathbf{e_{r}}.
\end{eqnarray}

A time-averaged description is valid when the frequency is high compared to the inverse of the viscous time scale \mbox{$(R_{2}-R_{1})^{2}/\nu$} which is true indeed.
Then the time- and frequency-dependent electric potential in eq.~\eqref{eq:Fdep_fin_cyl} is replaced by its effective value $V_{0}$.

In addition to the DEP force, a spherical particle undergoes viscous friction which is modelled by the Stokes’ law.
Neglecting the transient phase of the particle velocity, the DEP force is balanced by the drag force, which leads to a radial velocity dependent on the radial position only:
\begin{eqnarray}
 \mathbf{v_{p}} & = & \frac{F_{DEP}(r)}{6\pi\rho_{f}\nu R_{p}}\mathbf{e_{r}}.
\end{eqnarray}

For the used spherical particles we have an averaged size $R_{p}=5\,\mu$m, density $\rho_{p}=1100$ kg/m$^{3}$ and relative permittivity $\epsilon_{p}=4.6$.
They are suspended in silicone oil Elbesil B5 (physical properties, see table~\ref{table:propOil}) in a cylindrical gap with radii $R_{1}=5$ mm and $R_{2}=10$ mm.
Applying a typical temperature difference in the experiment of $\Delta T=7$ K and a typical electric potential of $7$ kV, the particle sedimentation velocity is $6.3\cdot10^{-3}$ mm/s at the inner cylinder where the electric gravity is the largest.
Here we used kinematic viscosity and density of the fluid at \mbox{$\vartheta\approx28.5$ $^{\circ}\rm{C}$}, the temperature at the warm inner cylinder (see table~\ref{table:tempdep}).  
We, therefore, assure that the particle sedimentation velocity due to the additional DEP force is negligible for PIV measurements.
Hence the tracer particles are suitable for use also in the high voltage a.c. environment to measure the fluid flow.

\section{Numerical model}\label{numerical-model}
\newcommand{\temp}{T}
The experimental situation described above can be modelled through the TEHD Boussinesq equations, which are based on the standard Boussinesq approximation for natural convection and augmented by DEP force and Gauss's law for describing the electric field inside the fluid as a function of temperature.
These equations for fluid velocity $\mathbf{v}$, pressure $p$, temperature $\temp$ and electric potential $\Phi$ are given by

\begin{eqnarray}\label{eq:num_model}
 \dt\mathbf{v} + (\mathbf{v}\cdot\nabla)\mathbf{v} - \nu\Delta\mathbf{v} + \frac{1}{\rho_{0}}\nabla p &=& \!\! \alpha_{E}(\nabla\Phi)^{2}\nabla \temp - \alpha g \mathbf{e_z} (\temp - T_0) \notag \\
 \nabla\cdot\mathbf{v} &=& 0 \notag \\
 \dt \temp + (\mathbf{v}\cdot\nabla) \temp - \kappa \Delta \temp &=& 0 \\
 -\nabla\cdot(\epsilon_{0}\epsilon_{r}[1-\gamma (\temp-T_0) ]\nabla\Phi) &=& 0 \notag,
\end{eqnarray}
with reference temperature $T_{0}$ and $\alpha_{E}=\epsilon_{0}\epsilon_{f}\gamma/(2\rho_{0})$. The corresponding boundary conditions
are given by
\begin{equation}
\begin{aligned}
\mathbf{v} &= 0 &&\text{ on } \Lambda \notag \\
\temp &= T_i &&\text{ on } \{r = R_i\}, \; i = 1,2 \notag \\
\nabla \temp \cdot \vec{n} &= 0 &&\text{ on } \{z = 0\} \cup \{ z = h \} \notag \\
\Phi &= \delta_{i2} V_0 &&\text{ on } \{r = R_i\}, \; i = 1,2 \notag \\
\nabla \Phi \cdot \vec{n} &= 0 &&\text{ on } \{z = 0\} \cup \{ z = h \}, \notag   \\
\end{aligned}
\end{equation}
with $\Lambda$ denoting the complete boundary (see also~\citet{Yoshikawa:2013}). The second term of the electrical body force~\eqref{eq:electric_force} enters  the momentum equation in~\eqref{eq:num_model}, whereas the third term in~\eqref{eq:electric_force} is contained in the generalised pressure \mbox{$p = p_{hyd} - \frac{1}{2} \rho_0 \left( \frac{\partial \epsilon}{\partial \rho} \right)_{T=T_0} \mathbf{E}^{2} \ $}.
 
Our method for approximately solving this system of partial differential equations is based on the \mbox{Finite Element Method (FEM)} for discretization in both space and time.
For the spatial discretization, ~\eqref{eq:num_model} is reformulated in cylindrical coordinates and continuous Lagrange finite elements are used on a hexahedral mesh. In order to obtain a stable discretization for the incompressible flow part of~\eqref{eq:num_model}, $(\mathbf{v},p)$ is approximated by means of the Taylor-Hood element $\mathbb{Q}_2 \times \mathbb{Q}_1$~\citep{Girault:2011}.
The $\mathbb{Q}_2$ element is used for both $\temp$ and $\Phi$. 
The temporal discretization is implemented by a Petrov-Galerkin formulation with continuous trial and discontinuous test functions~\citep{Schieweck:2010}. In this way, the resulting discretised problem can be addressed in a time-stepping manner, similar to the well-known Crank Nicolson method. There, the arising set of nonlinear algebraic equations for each time step is solved by the Newton-Raphson method. For solving the associated linear systems, the GMRES method~\citep{Saad:2003} is applied with block-wise incomplete LU factorization as a preconditioner. 

The implementation of the method is based on the open source FEM package HiFlow$^{3}$~\citep{Gawlok:2017}. 
Depending on the specific experimental configuration, the presented numerical results are obtained for meshes consisting of 96,000 to 768,000 hexahedrons, resulting in approximately $4 \cdot 10^{6}$ to $2.7 \cdot 10^{7}$ spatial degrees of freedom respectively. For the temporal discretization, the time domain $[0,\mathcal{T}]$ is split into sub-intervals of length $\tau = 0.05$ s or $\tau = 0.1$ s. Both spatial and temporal discretization parameters are chosen such that decreasing the average cell width and the time step size by a factor of 2 do not yield significantly different results. As the initial condition, we use the stationary solution for the natural convection case, i.e., $\alpha_{E}=0$ and $\dt\mathbf{v}, \dt \temp$ are neglected in~\eqref{eq:num_model}. 
Against the background of a transient solution converging towards a stationary state, the final time $\mathcal{T}$ of each simulation is chosen such that the residual of the stationary version of~\eqref{eq:num_model} is below a tolerance of $10^{-3}$ times the initial residual and significant change in the solution vector is no longer observed.
In the precise simulations considered on page 9 in figure~\ref{fig:gradient}(a) and ~\ref{fig:TNumSlice}, there holds $\mathcal{T} = 650$ s, whereas $\mathcal{T} = 250$ s for the data visualized in~\ref{fig:gradient}(b).

Nusselt numbers are computed by integrating the radial temperature gradient over a thin, vertical annulus and taking the average w.r.t. $r$, as an approximation to the surface integral over the inner wall,
\begin{eqnarray}\label{eq:Nu}
\Nu & = & \frac{ht(\temp)}{ht(\temp_{cond})}
\end{eqnarray}
with
\begin{eqnarray}
ht(\temp) & = & \frac{1}{0.05 d} \int_{r_1}^{r_1+0.05d} \int_0^{2 \pi} \int_0^{h} \frac{\partial}{\partial r} \temp (r, \phi, z) r {\rm d}(r, \phi, z). \notag
\end{eqnarray}
Here, $\temp_{cond}$ denotes the temperature field that is present in the pure conduction case, i.e. without DEP force and natural \mbox{gravity}.
Computing the finite element based heat transfer by means of a volume integral over a thin annulus turned out to yield more accurate results than an integral over the inner cylinder surface. 
For this investigation, we considered the case of periodic boundary conditions on top and bottom plate and compared both types of integrals with the (surface) heat transfer calculated from the known analytical solution, given e.g. in~\citet[][eq. (13)]{Yoshikawa:2013}. 
This might be due to the fact that volume integrals are typically less sensitive w.r.t. discretisation errors of the finite element scheme than surface integrals.

\section{Experimental procedure and post-processing}\label{protocol}
\begin{figure}
\centering
\includegraphics[width=0.45\textwidth]{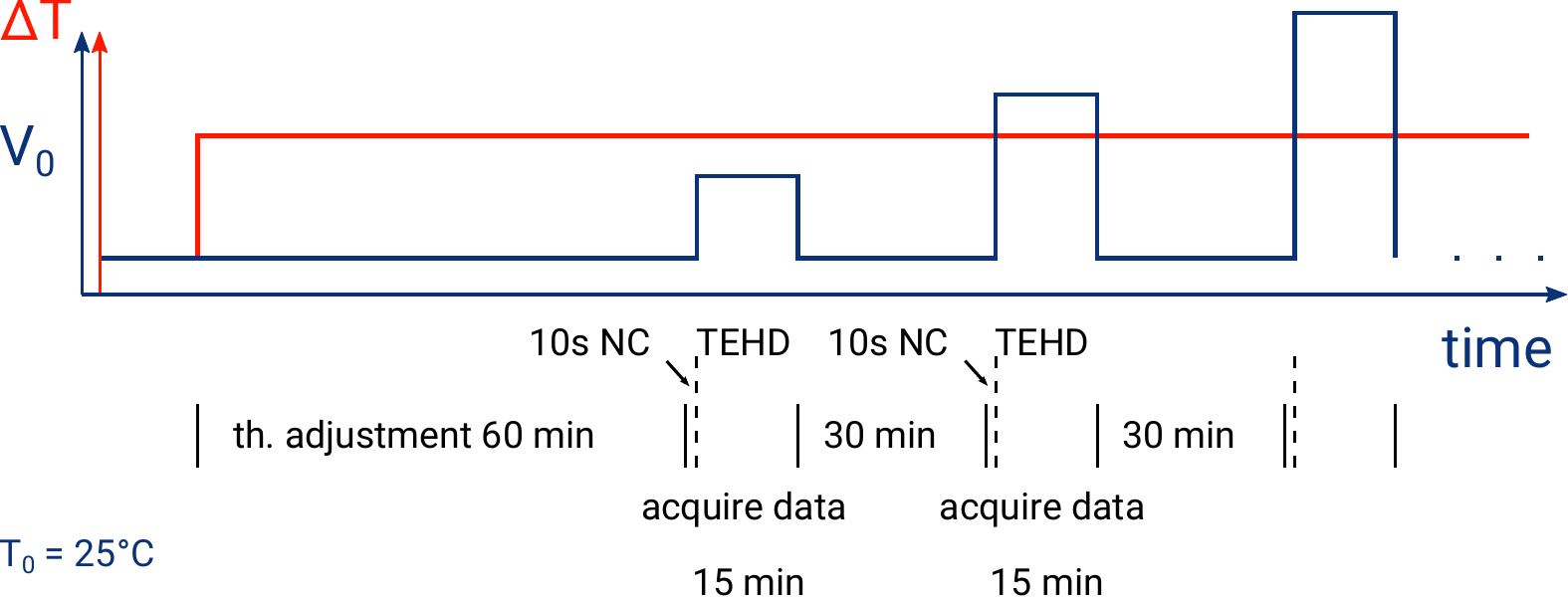}
\caption{Realization of experiments and data acquisition.}
\label{fig:protocol}
\end{figure}

\begin{figure*}
\centering
\includegraphics[width=0.79\textwidth]{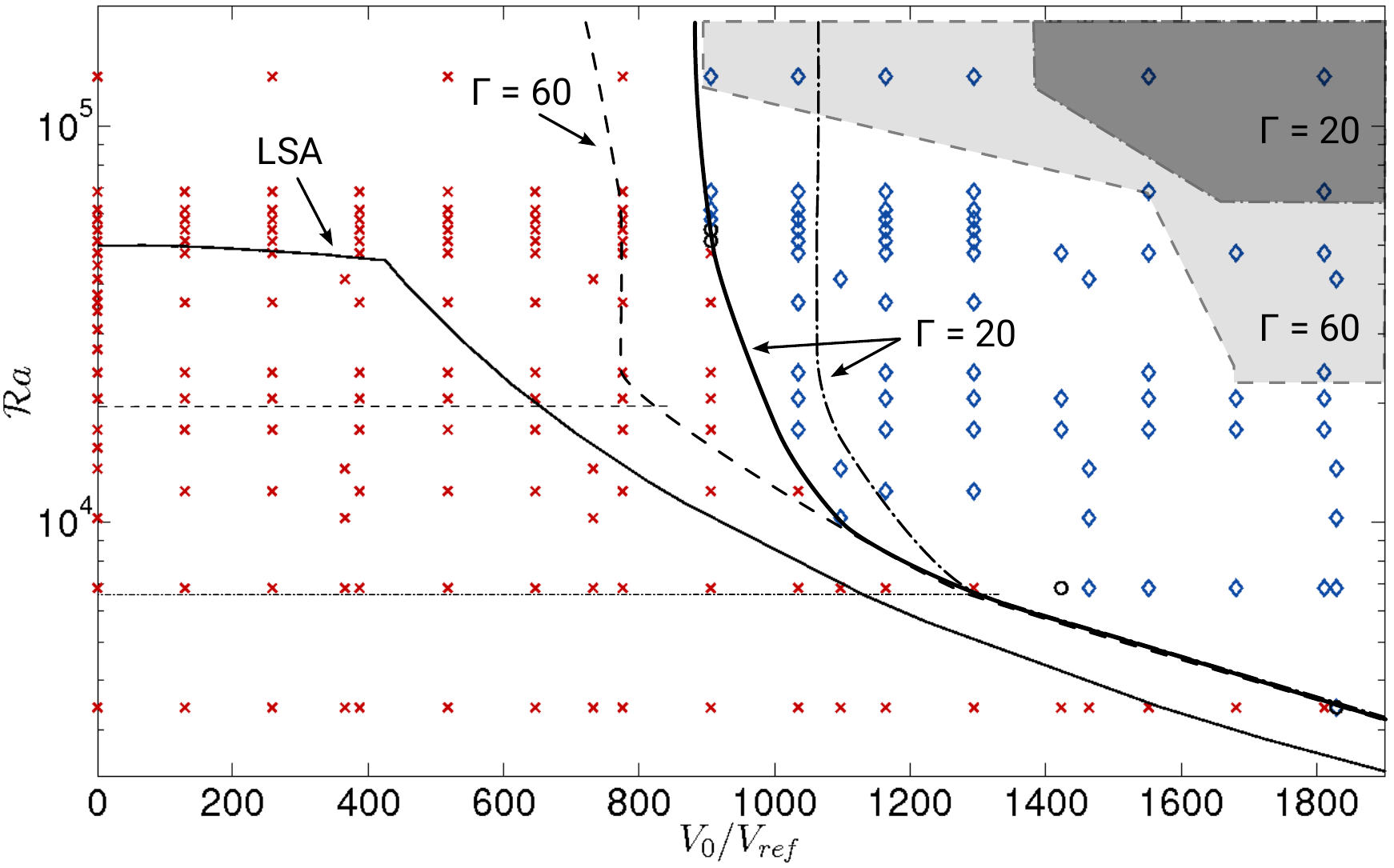}
\caption{Regime diagram spanned by $V_{0}/V_{ref}$ and thermal Rayleigh number \Ra. The symbols indicate Shadowgraph experiments
with $\Gamma = 20$. Red crosses and black circles stand for unicellular flow and indifferent flow. Blue diamonds outside/inside the dark-grey region stand for stationary/time-dependent columnar vortices. The thin solid line shows the stability diagram obtained by linear
stability analysis ($\Gamma = \infty$,~\cite{Meyer:2017a}). The thick solid and the dot-dashed line is the line of marginal stability obtained for the shadowgraph and PIV measurements ($\Gamma = 20$), respectively. The dashed line is the experimental transition line for $\Gamma = 60$ (PIV). The dark- and light-grey regions indicate experiments where oscillatory modes have been observed for $\Gamma = 20$ and $\Gamma = 60$, respectively. The two horizontal lines visualise the criterion found by~\citet[][eq. (9) and (10)]{Lopez:2015} for the transition from conductive to convective regime of the unicellular flow for $\Gamma = 20$ (dot-dashed line: \Ra$_{C}=6587$) and for $\Gamma = 60$ (dashed line: \Ra$_{C} =1.963 \cdot 10^{4}$).}
\label{fig:regime}
\end{figure*}

Each experimental day began in an isothermal state at room temperature $21-22^{\circ}\rm{C}$.
We set the cylindrical walls to its operating temperature needed to achieve the appropriate temperature difference $\Delta T=T_{1}-T_{2}$ between the outer wall of the inner cylinder and inner wall of the outer cylinder maintaining the reference temperature at $T_{0}=25^{\circ}\rm{C}$ (q.v. figure~\ref{fig:protocol}).
We kept the temperature difference constant for the whole day. 
Due to different heat conductivity of the cylinders in case of PIV measurements we took thermal losses into account.
The inner cylinder is made of aluminium, an excellent heat conductor ($\lambda_{1}$ at 25$^{\circ}\rm{C}\approx235$ W/(mK)).
With its thickness of 2 mm the thermal resistance is very low and negligible, therefore we applied no correction.
The outer cylinder is made of borosilicate glass, a material with a high thermal resistance ($\lambda_{2}$ at 20$^{\circ}\rm{C}\approx1.2$ W/(mK)).
For calculation of the corrected temperature at the outer wall of the outer cylinder, we used the formula given in~\citet[][eq. 5]{Futterer_Dahley:2016}.
They derived the equation from Fourier's law for heat conduction. 
In modified form it reads
\begin{eqnarray}
 T_{2,o} & = & (T_{2}-T_{1})\frac{\lambda d}{\lambda_{2}d_{2}} + T_{2,i},
\end{eqnarray}

where $T_{2,o}$ is the absolute temperature at the outer wall of the outer cylinder and $d_{2}=2$ mm its thickness.
We waited 1 hour to attain a thermally well-balanced state.
Afterwards, we started recording data.
The first 10 seconds at the base state of natural convection.
Then, we switched on the voltage.
An experiment ended after 15 minutes switching of high voltage followed by a resting time of 30 minutes where the system attains back to its thermally balanced state.
We repeated the procedure whereat the voltage was increased stepwise from one experiment to the next.
The thermal and electric Rayleigh number \Ra$\,$ and $L$, namely
\begin{eqnarray}
 \mathcal{R}a = \frac{\alpha g \Delta T d^{3}}{\nu\kappa} & \text{and} & L = \frac{\alpha g_{e} \Delta T d^{3}}{\nu\kappa},
\end{eqnarray}
with electric gravity of the base state~\citep{Mutabazi:2016}
\begin{eqnarray}\label{eq:eg}
 g_{e} & = & F(\gamma\Delta T,\eta,r)\frac{\epsilon_{0}\epsilon_{f}\gamma}{\rho\alpha r^{3}}\left(\frac{V_{0}}{\ln(\eta)}\right)^{2}
\end{eqnarray}
and
\begin{eqnarray}
 F(\gamma\Delta T,\eta,r) & = & \left[\frac{\gamma\Delta T}{\ln(1-\gamma\Delta T)}\right]^{2}\left[1-\gamma\Delta T\left(\frac{\ln(r/R_{2})+1}{\ln{\eta}}\right)\right]\times \notag \\
 & & \times\left[1-\gamma\Delta T\frac{\ln(r/R_{2})}{\ln(\eta)}\right]^{-3}
\end{eqnarray}
are calculated by considering the fluid properties at the reference temperature (see table~\ref{table:tempdep}).
In the following, we refer the electric Rayleigh number to the midgap radius \mbox{$r = 7.5$ mm}.
For each tuple ($\Delta T$,$V_{0}$) a Hovm{\o{}}ller-diagram from PIV at midgap \mbox{($r=7.5$ mm)} has been estimated.
Similar diagrams for shadowgraph measurements have been estimated either fixing the angle $\varphi$ or the radius $r=7.5$ mm.
Ernest Aabo Hovm{\o{}}ller (1912--2008), a Danish meteorologist, first introduced in 1949 this kind of diagram which shows data in a space-time plot~\citep{Hovmoeller:1949}.
In order to prove whether the flow is axially aligned and columnar, we calculated the radial average of the axial gradient of velocity at the final time where the flow is stationary
\begin{eqnarray}
 \left\langle \frac{\partial}{\partial z} {\rm sgn}(v_{z}) \parallel v(r,z) \parallel \right\rangle_{r} (z) & = & \frac{\int_{r} \frac{\partial}{\partial z} {sgn}(v_{z})(v_{r}^{2}+v_{z}^{2})^{1/2} {\rm d}r}{\int_{r}{\rm d}r}.
\end{eqnarray}

Additionally, in order to obtain information regarding the onset of instability, we estimated at midheight the radial average of the radial velocity component
\begin{eqnarray}
 \langle v_{r}(h/2) \rangle (t) & = & \frac{\int_{r} v(r,z=h/2,t) {\rm d}r}{\int_{r}{\rm d}r}.
\end{eqnarray}

\section{Results}\label{results}

\begin{figure*}
\centering
\includegraphics[width=\textwidth]{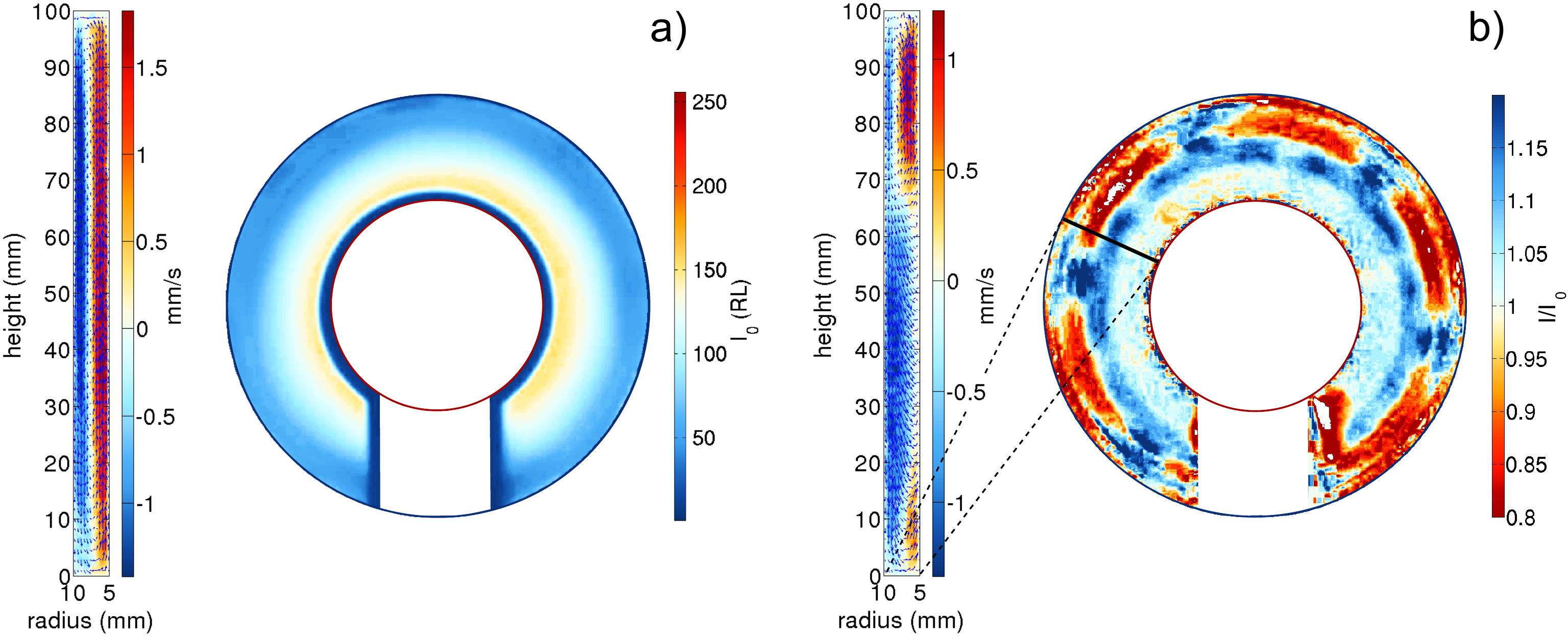}
\caption{Composition of PIV and shadowgraph measurement (left and right in each subfigure (a),(b)). The flow in the cavity with aspect ratio $\Gamma=20$ is caused by a temperature difference $\Delta T=7$ K (\Ra$\;=23946$). 
Shadowgraph images show the top view of (a) the reference light intensity $I_{0}$ (red channel) and (b) the normalised light intensity $I/I_{0}$ of the axially integrated flow in the radial-azimuthal plane. Blue/red colour refers to denser/lighter fluid.
PIV images show the velocity field in the meridional plane. Blue/red colour refers to down/upward flow. 
(a) Flow without electric potential $V_{0}$. The velocity field shows the typical base flow for natural convection (azimuthal wavenumber $n = 0$ in shadowgraph image).
(b) Flow with electric potential $V_{0}=7$ kV ($L = 15220$). The flow undergoes a transition to a stationary and axially aligned columnar structure (velocity field) of azimuthal wavenumber $n = 6$ (shadowgraph image). The black solid line refers to the plane for measurements in figure~\ref{fig:Hov-snap}.}
\label{fig:comp_dT7K}
\end{figure*}

An overview of the experiments is shown in figure~\ref{fig:regime}.
We chose different values of cylinder height $h$, temperature difference $\Delta T$ and electric potential $V_{0}$.
This gives a range of non-dimensional parameters $\Gamma$, \Ra$\,$ and $V_{0}/V_{ref}$.
For non-dimensionalisation of electric potential we used the reference electric potential $V_{ref}=(\rho\kappa\nu/\epsilon_{0}\epsilon_{f})^{1/2}$ suggested by~\citet{Yoshikawa:2013}.
The regime diagram is spanned by \Ra, $V_{0}/V_{ref}$ but it should be kept in mind that aspect ratio $\Gamma$, radius ratio $\eta$ and Prandtl number \Pra$\,$ may also vary.
For our investigation $\eta$ and \Pra$\,$ have been fixed.
Experiments with shadowgraph technique applied (aspect ratio $\Gamma=20$) have been marked with crosses, diamonds and circles. 
Red crosses indicate stability where unicellular laminar (natural) free convection dominates the flow.
Blue diamonds indicate thermal electro-hydrodynamic convection, where we found a pattern with azimuthal wavenumber $n\neq0$.
Black circles stand for indifferent flows.
Based on shadowgraph measurements we delineated the line of marginal stability (thick solid line) separating stable from the unstable flow.
Due to the fact that shadowgraph shows global variations of the flow we validate/substantiate those results with PIV measurements. 
In case of aspect ratio $\Gamma=20$ we found the line of marginal stability plotted as dot-dashed line and in case of aspect ratio $\Gamma=60$ as the dashed line. 
The thin solid line was computed by using a linear theoretical model used before in~\citet{Meyer:2017a}.
This model solves a set of linearized equations, namely the continuity equation, the Boussinesq approximated momentum equation, the energy equation and the Gauss' law of electricity.
The model has been adapted to the experimental geometry, i.e. it has a rigid inner and outer cylinder but it is considered to be infinite in the axial direction and hence neglects top and bottom lids.

For \Ra$\,<6800$ transition from stable unicellular laminar free convective flow to thermal electro-hydrodynamic convection is independent on aspect ratio.
The theoretically predicted line of marginal stability is validated reasonably good.
For \Ra$\;>6800$ and considering aspect ratio \mbox{$\Gamma=20$} we observe a slight offset of marginal stability regarding the measurement techniques (thick solid and dot-dashed line).
In case of $\Gamma=20$ the line of marginal stability found with shadowgraph and PIV show a slight offset.
The transition found with PIV lags behind shadowgraph technique.
The material used for the outer cylinder could be a source of the slight difference in the measured threshold of the transition.
Also, the azimuthal position of the laser plane plays an important role, in order to detect the radial velocity with an amplitude high enough.

Next, we consider destabilisation of the flow regarding aspect ratio (figure~\ref{fig:regime}, $\Gamma=20$ - dot-dashed line and $\Gamma=60$ - dashed line).
A comparison of the lines of marginal stability which are obtained with PIV shows, when \Ra$\;>6800$, the larger the aspect ratio is, the more unstable the flow is.
In the case of aspect ratio $\Gamma=60$, the line of marginal stability tends towards theoretical prediction for a cylinder with infinite axial direction.
It substantiates the transition to be linear in an interval ($V_{0}/V_{ref}$,\Ra) where the Boussinesq approximation is valid (i.e. for \Ra$\,\lessapprox 2.5\cdot10^{4} \equiv \Delta T\lessapprox 7.3$ K). 
For \Ra$\,$ high enough non-linear processes have to be considered which influence also the transition from stable to unstable flow.

Additionally, we draw the region where the flow undergoes a second transition and becomes time-dependent, e.g. the vertical wavenumber $k>0$ and the vertical frequency component \mbox{$\omega_{k}\neq0$}, for aspect ratio $\Gamma=20$ (light-grey) and aspect ratio $\Gamma=60$ (dark-grey).
But however, this will not be discussed.

\subsection{Regimes}\label{regimes}
\begin{figure}[t]
\centering
\includegraphics[width=0.43\textwidth]{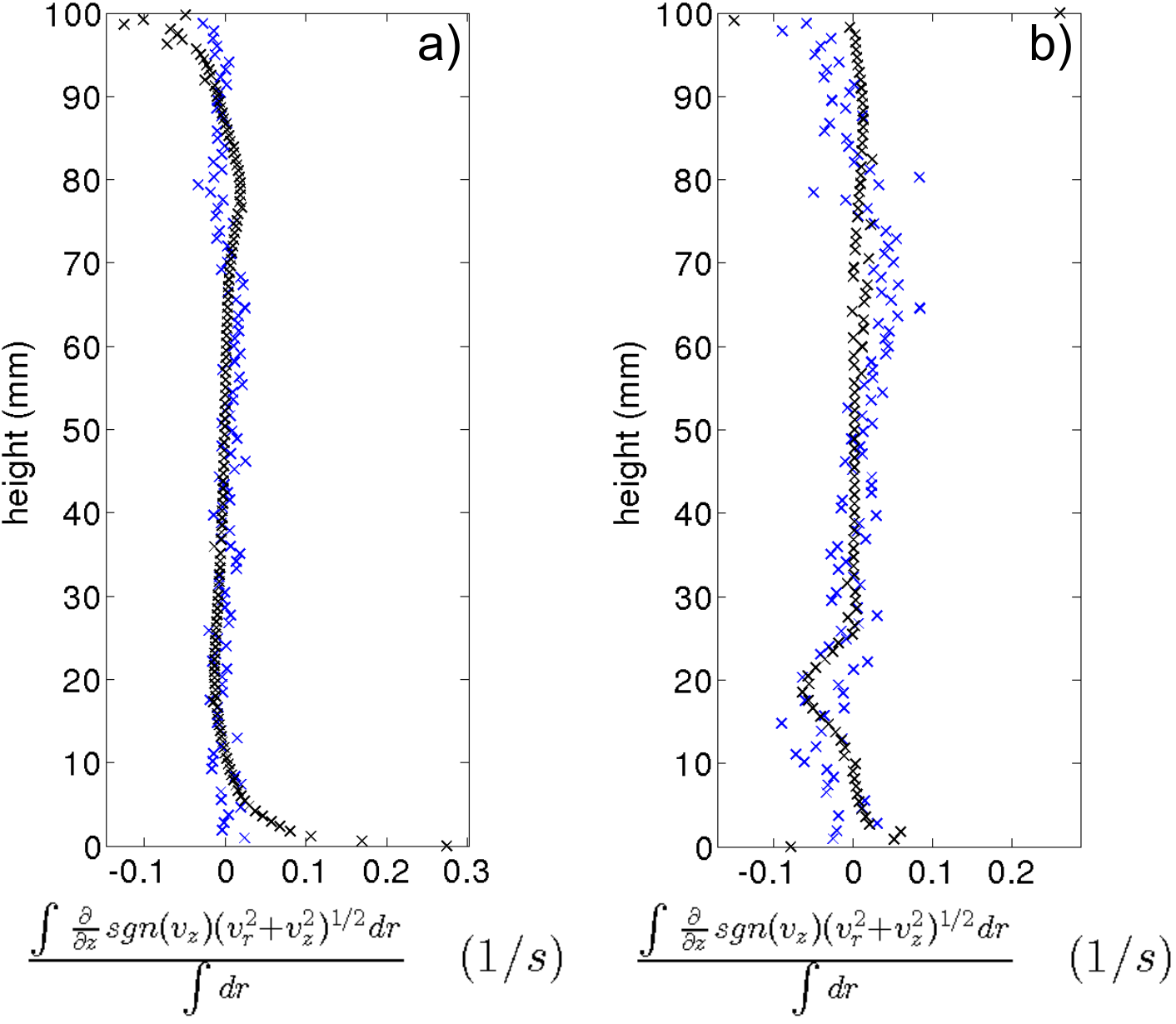}
\caption{Radial average of the axial gradient of velocity at the final time where the flow is stationary. $\Delta T = 2$ K \mbox{(\Ra$\;=6842$)}, $V_{0} = 6$ kV ($L=3173$) (a) and $\Delta T = 7$ K (\Ra$\;=23946$), \mbox{$V_{0} = 7$ kV} ($L=15220$) (b). Blue and black symbols correspond to the experiment and numerical simulation, respectively.}
\label{fig:gradient}
\end{figure}

As a first step, we investigated the flow with PIV caused by a radial temperature difference without electric potential.
\citet{Elder:1965a} found the critical thermal Rayleigh number of \Ra$_{C}\,\approx3\cdot10^{5}\pm30\%$ where unicellular laminar (natural) free convection in a vertical slot undergoes a transition to a stationary secondary flow.
A second set of streamlines appear, called cat-eye pattern, with one short vortice at the lower region and one large vortex that reaches the upper-end region. 
The uncertainty of \Ra$_{C}$ is large caused by the difficulty to detect the onset of the very weak secondary flow, especially when the vertical wavelength is large.
For the used apparatus with aspect ratio $\Gamma=19$, he found \Ra$_{C}=3.6\cdot10^{5}$ which is related to a radial temperature difference $\Delta T=27\pm2$ K. 
The critical Rayleigh number implies for our setup with $\Gamma=20$ and $60$ a radial temperature difference \mbox{$\Delta T\approx88$ K}.
The destabilising effect of radius ratio $\eta$ for high Prandtl numbers is very small~\citep[see][fig. 6]{Choi:1980}.
Therefore we conclude, a maximal applicable Rayleigh number of $2.053\cdot10^{5}$ ($\Delta T=60$ K) to both of the experiment cells is too small to find the cat-eye pattern.
\citet{Meyer:2017a} (fig.~\ref{fig:regime}, thin solid line) found with linear stability analysis in case of absence of electric potential critical modes in the form of oscillatory axisymmetric vortices (vertical wavenumber $k_{C}\approx2.5$) of the thermal instability (see also \citet{Bahloul:2000}).
In fact and contrary to previous literature, we did not observe a stationary secondary flow superposed on the unicellular base flow.
Hence, the initial condition and base flow we start with is the unicellular laminar (natural) free convection.
Figure~\ref{fig:comp_dT7K}(a) shows the base flow measured with PIV (left) and shadowgraph (right).
Obviously, in PIV the unicellular pattern can be observed.
Down/upward flow (denser/lighter fluid) is represented in blue/red colour.
In the shadowgraph image, the false colour representation of the reference light intensity $I_{0}$ shows no variation in azimuthal direction (wavenumber $n=0$) and a linear decrease of light intensity in the radial direction, whereat blue/red colour refers to denser/lighter fluid.
Telecentric light rays emitted in close proximity to the warm inner cylinder are refracted towards regions of cooler fluid inducing a region of 'no light' (dark blue region in close proximity to the warm inner cylinder).

Next, we study the flow with an electric potential applied.
Exemplarily, we consider the flow caused by a temperature difference \mbox{$\Delta T= 7$ K} (\Ra$\;=23946$) and an electric potential \mbox{$V_{0} = 7$ kV} ($L = 15220$) (figure~\ref{fig:comp_dT7K}(b)).
Immediately after switching on electric potential convective instability sets in.
Warmer fluid adjusts to regions with the less intense electric field and cooler fluid to regions with the more intense electric field.
The transient state converges to a stationary state, where the flow saturates/establishes.
In the radial-azimuthal plane, the instability forms convective plumes with azimuthally alternating radially in/outward directed jets.
The jets transport cool/warm fluid.
The plumes are equidistantly distributed showing an azimuthal wavenumber $n=6$.
Along the z-axis, there is no axial gradient of velocity (figure~\ref{fig:gradient}(b)), except in close vicinity to the end-plates where the boundary layer circulation dominates.
Therefore, we argue the plumes are axially aligned and show a stationary columnar structure 
trapped between upper and lower boundary layer circulations.
The vertical extent of the upper and lower boundary layer circulation depends on temperature difference, electric potential and aspect ratio.
A stabilising effect due to adiabatic upper and lower boundaries was previously found by \citet{GillDavey:1969} for natural convection between vertical concentric cylinders when \Ra$\;>300\Gamma$~\citep[cf. also][]{Choi:1980}.
This fact reflects also in TEHD flow.
Generally, it may be said (if \Pra$\;=const.$), the lower the aspect ratio of the experiment cell the larger the vertical extent of the boundary layer and the higher its stabilising effect on the flow (see e.g. figure~\ref{fig:regime}).

\begin{figure}[t]
\centering
\includegraphics[width=0.5\textwidth]{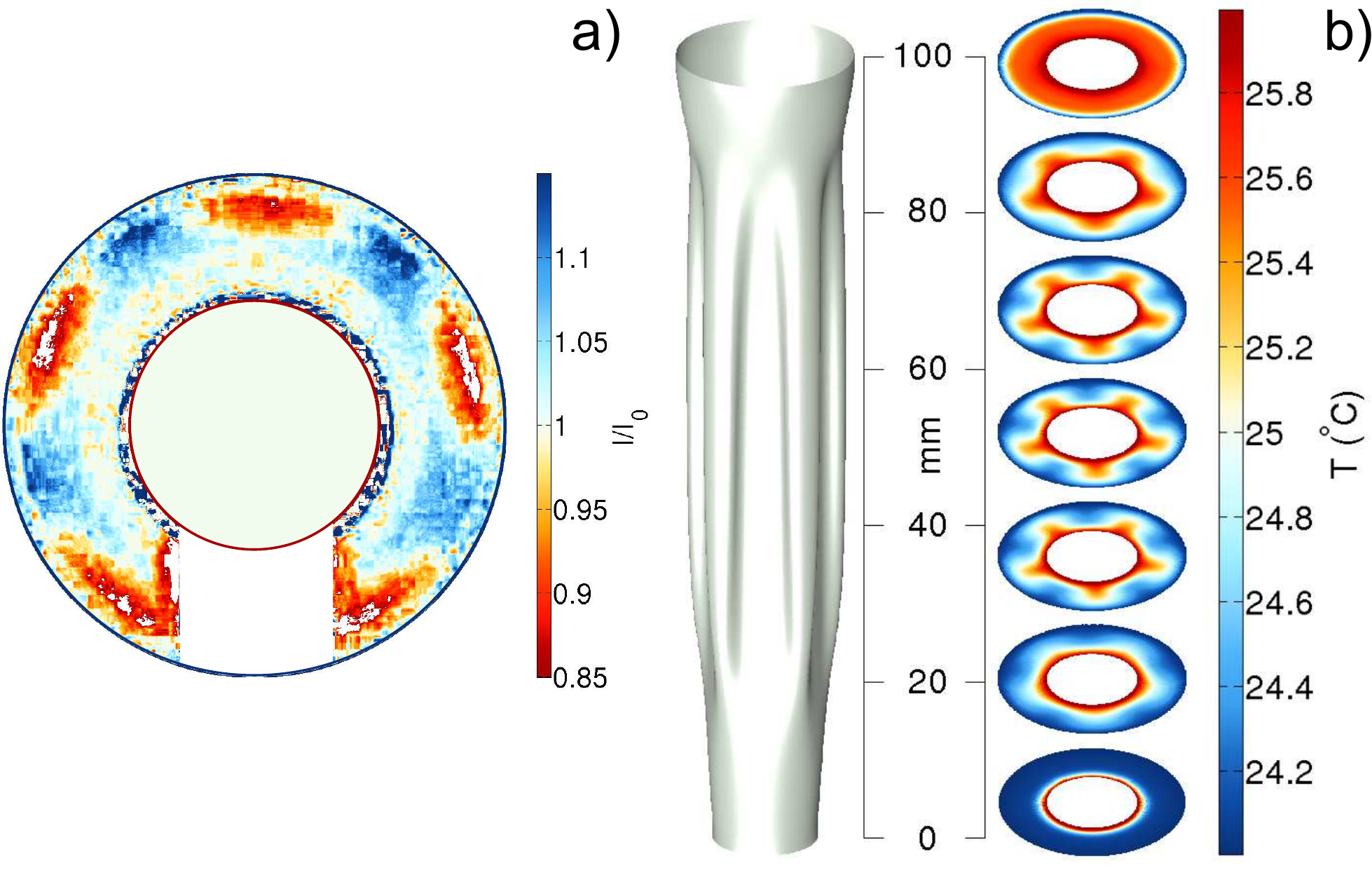}
\caption{Flow caused by a temperature difference $\Delta T=2$ K (\Ra$\;=6842$) and an electric potential \mbox{$V_{0}=6$ kV} ($L = 3173$). Experiment: Shadowgraph measurement (a). Numerical simulation: isosurface of the background temperature $T_{0}=25^{\circ}$C (middle) and 3D temperature distribution (b).}
\label{fig:TNumSlice}
\end{figure}

\begin{figure*}
\centering
\includegraphics[width=0.98\textwidth]{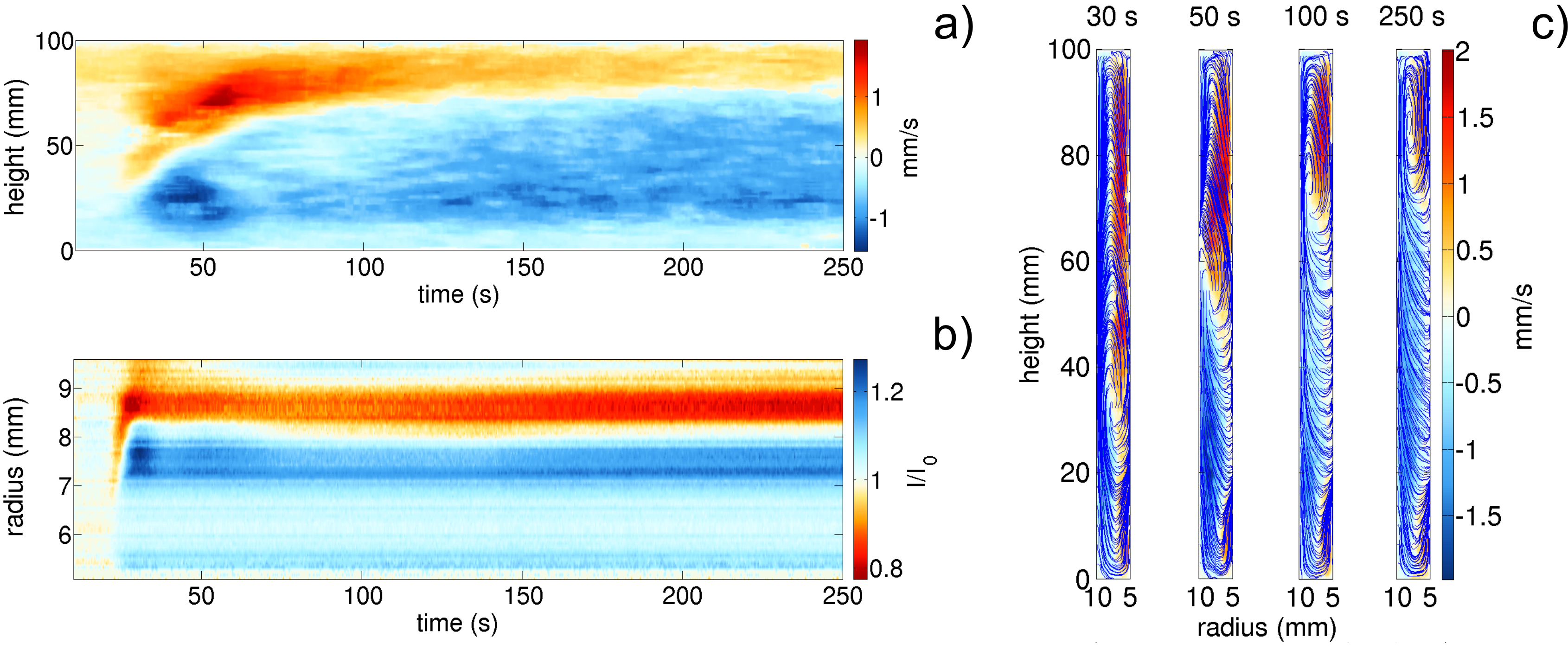}
\caption{Experiments with a temperature difference \mbox{$\Delta T=7$ K} (\Ra$\;=23946$) and electric potential $V_{0}=7$ kV ($L = 15220$) in the cavity with aspect ratio $\Gamma=20$ (cf. figure~\ref{fig:comp_dT7K}(b)). PIV measurement is performed in a plane along the black line in figure~\ref{fig:comp_dT7K}(b). 
(a) Hovm{\o{}}ller-diagram spanned by height and time shows $\rm{sgn}(v_{z})\parallel \mathbf{v} \parallel$ at midgap $r=7.5$ mm. Colour same as in PIV image in figure~\ref{fig:comp_dT7K}(b). The instability occurs first in the lower part of the gap and develops upward.
(b) Hovm{\o{}}ller-diagram spanned by radius and time shows normalised light intensity $I_{0}$ (along black line fig.~\ref{fig:comp_dT7K}(b)). Blue/red colour refers to denser/lighter fluid. Instability sets in almost at same time like in PIV measurement and shows density/temperature inversion.
(c) Snapshots of $\rm{sgn}(v_{z})\parallel \mathbf{v} \parallel$ (colour) in a plane spanned by radius and height superposed by associated streamlines at $t=[30,\,50,\,100,\,250]$ s. Colour same as in (a).}
\label{fig:Hov-snap}
\end{figure*}

Figure~\ref{fig:TNumSlice} shows another example of axially aligned convective plumes having a stationary columnar structure trapped between upper and lower boundary layer circulations.
The flow caused by a temperature difference \mbox{$\Delta T= 2$ K} (\Ra$\;=6842$) and an electric potential \mbox{$V_{0} = 6$ kV} ($L = 3173$) develops in close proximity to the line of marginal stability.
The experimental shadowgraph image (figure~\ref{fig:TNumSlice}(a)) shows the top view of light intensity $I/I_{0}$.
We identify equidistantly distributed convective plumes with azimuthal wavenumber $n=5$.
Nevertheless, due to axially integrated light intensity, it is speculative to assume a columnar structure of the convective plumes although the axial gradient of velocity, for the most part, vanishes (figure~\ref{fig:gradient}(a)).
Numerical simulation substantiates our hypothesis. A solution converging towards a stationary state is shown in~figure~\ref{fig:TNumSlice}(b).
The 3D isosurface of the background temperature $T_{0}=25^{\circ}$C (figure~\ref{fig:TNumSlice}(b), left) and the 3D temperature distribution (figure~\ref{fig:TNumSlice}(b), right) apparently shows the stationary columnar nature of the axially aligned convective plumes.
Indeed, we found the stationary columnar structure of the convective plumes for numerical simulations covering an area of tuples ($\Delta T$,$V_{0}$) ranging from (2 K, 0 kV) to (20 K, 7 kV).

\subsection{Pattern formation and growth}\label{growth}
After the first transition, in the interval of ($V_{0}/V_{ref}$,\Ra) where the flow is columnar and stationary, the flow measured in the azimuthally fixed PIV-plane destabilises and evolves always in the same manner.
Exemplary, we study in more detail the flow caused by a temperature difference \mbox{$\Delta T=7$ K} (\Ra$\;=23946$) and electric potential $V_{0}=7$ kV ($L = 15220$) in the cavity with aspect ratio $\Gamma=20$ (black line fig.~\ref{fig:comp_dT7K}(b)).
When the electric potential is switched on the electric gravity of the base state $g_{e}$ (cf. eq.~\ref{eq:eg}) is normal to Earth's gravity $g$ and proportional to 1/r$^{3}$.
If we refer the radius to midgap then dielectrophoretic acceleration has a value of 0.6356$g$.
The resultant effective gravity has a value of 1.1849$g$ and Earth's gravity and effective gravity draw an angle of 32.44$^{\circ}$.
Figure~\ref{fig:Hov-snap} show Hovm{\o{}}ller-diagrams obtained from two independent measurements, one performed with PIV and the other with the shadowgraph technique applied.
The first, spanned by height and time (figure~\ref{fig:Hov-snap}(a)) shows $\rm{sgn}(v_{z})\parallel \mathbf{v} \parallel$ at midgap $r=7.5$ mm.
Colour is the same as the PIV image in figure~\ref{fig:comp_dT7K}(b).
\begin{figure}
\centering
\includegraphics[width=0.48\textwidth]{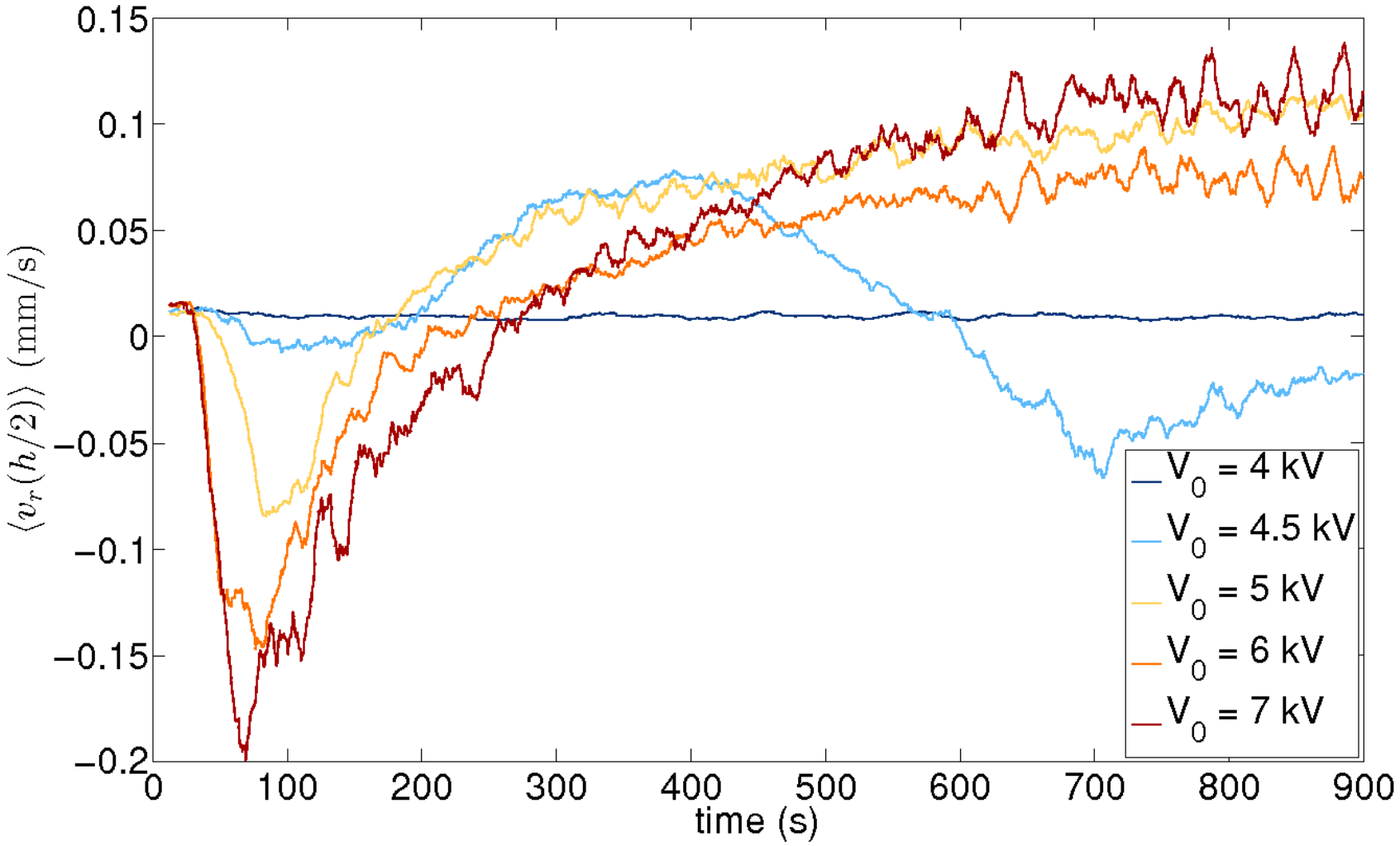}
\caption{Radial velocity at midheight averaged over gap as function of time. A property we used to define quantitative if the flow is stable/unstable. We applied a temperature difference of \mbox{$\Delta T=7$ K} (\Ra$\;=23946$) in the cavity of aspect ratio $\Gamma=20$ and varied electric potential.}
\label{fig:Stable_or_not}
\end{figure}
The second (figure~\ref{fig:Hov-snap}(b)), spanned by radius and time shows normalised light intensity $I/I_{0}$.
Blue/red colour refers to denser/lighter fluid.
Additionally, figure~\ref{fig:Hov-snap}(c) shows snapshots of $\rm{sgn}(v_{z})\parallel \mathbf{v} \parallel$ (colour same as in figure~\ref{fig:Hov-snap}(a)) and associated streamlines at $t=[30,\,50,\,100,\,250]$ s.
After electric potential is switched on it takes some seconds to destabilise the base flow.
Destabilisation starts first in the lower and midheight part (see also snapshot at $t=30$ s).
The interface separating up/downward directed base flow becomes wavy, whereat more pronounced amplitude is visible in the lower part of the gap.
Obviously, the lower part wins the competition and instability completely evolves, down here (see also snapshot at $t=50$ s).
During the transient state, the structure grows upward (see also snapshot at $t=100$ s) and reaches its maximal vertical extent when the columnar plume-like instability equals the state of equilibrium (see also snapshot at $t=250$ s).
Nevertheless, thin boundary layers remain at vertical sidewalls.
The evolution of the pattern in shadowgraph (fig.~\ref{fig:Hov-snap}(b)) is visible as density/temperature inversion.

\begin{figure}
\centering
\includegraphics[width=0.475\textwidth]{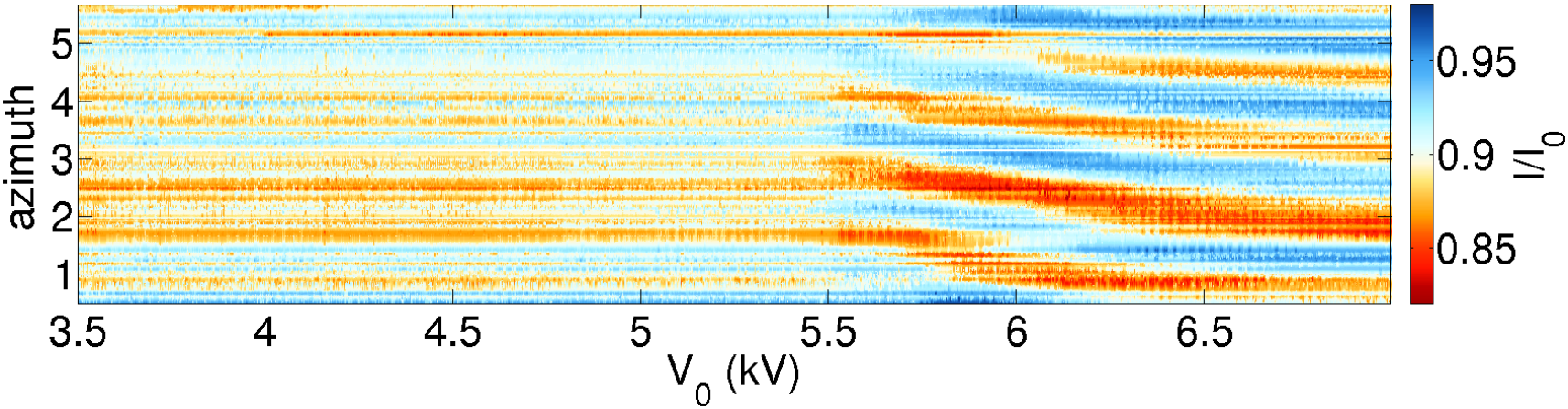}
\caption{Hovm{\o{}}ller-diagram at midgap ($r=7.5$ mm) of the flow measured with shadowgraph technique. The flow is caused by a temperature difference \mbox{$\Delta T=2$ K} (\Ra$\,=6842$) in the cavity of aspect ratio $\Gamma=20$. Electric potential is continuously increased over 120 minutes from $V_{0} = 3.5$ kV ($L=1080$) to $V_{0} = 7$ kV ($L=4319$). Blue/red colour refers to denser/lighter fluid. The flow undergoes  a transition to a stationary and axially aligned columnar structure of azimuthal wavenumber $n = 5$ (see also fig.~\ref{fig:TNumSlice}) at $V_{0}\approx5.5$ kV ($L=2667$). The azimuthal position of the columnar structure strongly depends on electric potential. An increase of electric potential shifts the pattern counterclockwise.}
\label{fig:ramp}
\end{figure}

The growth of perturbation might also be considered by looking on the temporal behaviour of the radial velocity at midheight averaged over gap (figure~\ref{fig:Stable_or_not}), being negative/positive is referred to radially outward/inward velocity. 
Also, we used the time-series of $\langle v_{r}(h/2) \rangle$ to define more quantitative if the flow is stable/unstable (cf. figure~\ref{fig:regime}).
In case of stable base flow there is no radial velocity, cf. flow with electric potential \mbox{$V_{0}=4$ kV} ($L=4970$) applied.
Increasing the electric potential, such as $V_{0}=4.5$ kV ($L=6290$), we easily distinguish the flow from being stable.

Dependent on the Rayleigh number we investigated the transition from stable unicellular laminar free convective flow to thermal electro-hydrodynamic convection in more detail with the shadowgraph technique (cf. also fig.~\ref{fig:regime} thick solid line).
Again, we exemplary applied a temperature difference of \mbox{$\Delta T=2$ K} (\Ra$\,=6842$).
In order to achieve a quasi-stationary variation of electric potential, it was continuously increased over 120 minutes from $V_{0} = 3.5$ kV ($L=1080$) to $V_{0} = 7$ kV ($L=4319$).
Figure~\ref{fig:ramp} shows the normalised light intensity $I/I_{0}$ in a Hovm{\o{}}ller-diagram at midgap ($r=7.5$ mm) spanned by azimuth and electric potential.
Blue/red colour refers to denser/lighter fluid.
The flow undergoes a transition to thermal electro-hydrodynamic convection with azimuthal wavenumber $n = 5$ (see also fig.~\ref{fig:TNumSlice}) at $V_{0}\approx5.5$ kV ($L=2667$).
Interestingly, the azimuthal position of the columnar structure depends strongly on electric potential.
An increase in electric potential shifts the pattern counterclockwise.
If electric potential and therefore electric gravity is high enough the pattern gets arrested in its azimuthal position.
The azimuthal shift is approximately $\pi/2$.
It depends on \Ra, the azimuthal shift decreases rapidly with increasing \Ra$\,$ (not shown).

We considered also the case of inverse quasi-stationary variation of electric potential, hereby it was continuously decreased over 120 minutes from $V_{0} = 7$ kV ($L=4319$) to $V_{0} = 3.5$ kV ($L=1080$).
On startup such as $V_{0} = 7$ kV ($L=4319$), we found the pattern on the same azimuthal position than compared to the experiment with increasing electric potential. 
Contrary, a decrease of electric potential shifts the pattern clockwise over a range of $V_{0}\approx 1-1.5$ kV ending in an azimuthally arrested pattern before the transition to stable unicellular laminar free convective base flow (here not shown).

\begin{figure}
\centering
\includegraphics[width=0.49\textwidth]{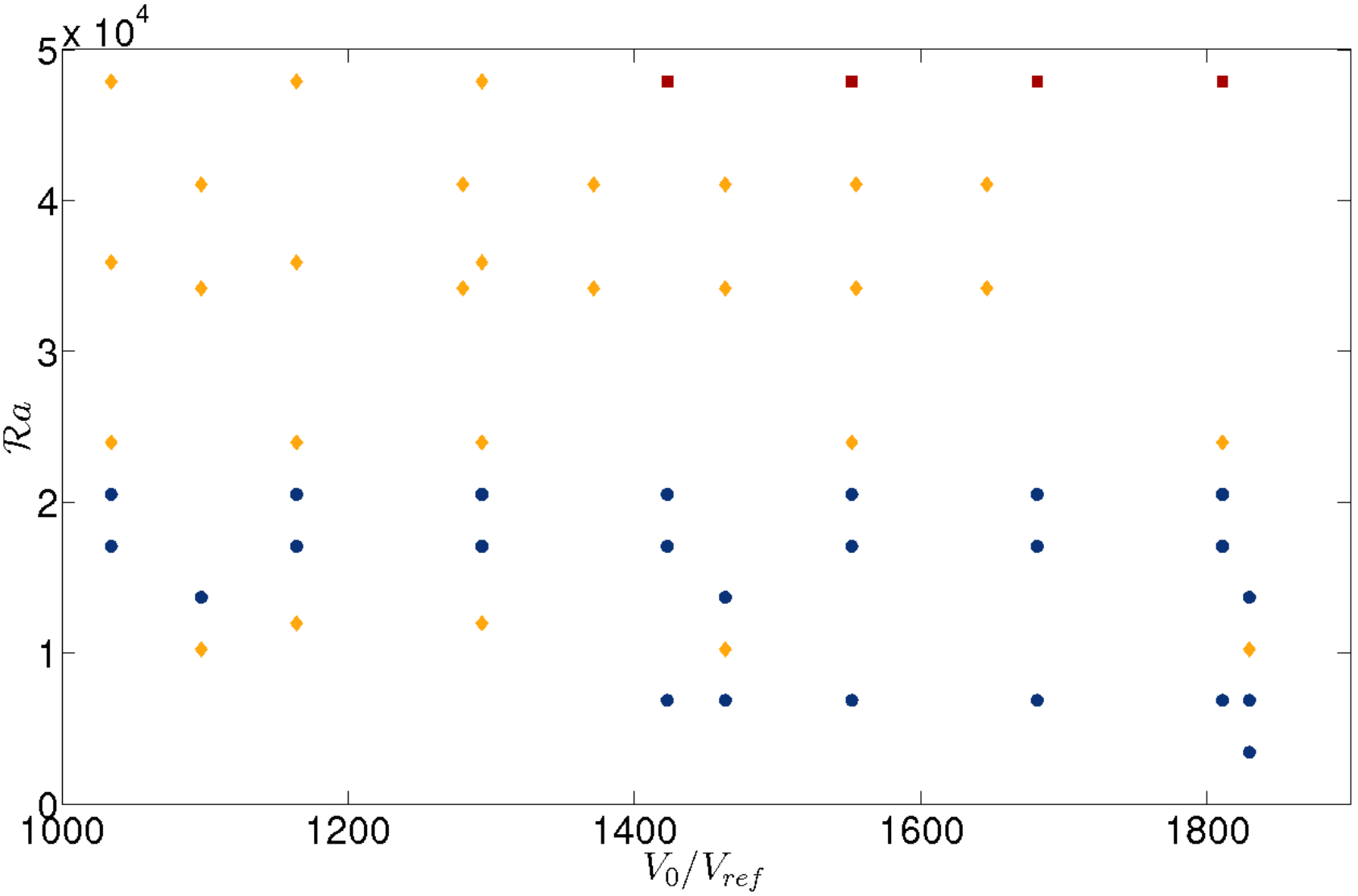}
\caption{Diagram spanned by dimensionless electric potential $V_{0}/V_{ref}$ and thermal Rayleigh number. It shows the azimuthal wavenumber $n$ estimated from shadowgraph measurements in the cavity with aspect ratio $\Gamma=20$. Blue circles: $n=5$; yellow diamonds $n=6$, red squares $n=7$.}
\label{fig:wavenumber}
\end{figure}

In the range of ($V_{0}/V_{ref}$,\Ra) after the first transition, where the flow is columnar and stationary, we identified with shadowgraph technique the azimuthal wavenumber $n$ of the instability.
We draw this quantity in a diagram, spanned by dimensionless electric potential $V_{0}/V_{ref}$ and thermal Rayleigh number.
Figure~\ref{fig:wavenumber} shows flow with azimuthal wavenumber $n=5$ as blue circles, $n=6$ as yellow diamonds and $n=7$ as red squares, respectively.
Obviously, there is no change in wavenumber when electric potential increases up to $V_{0}/V_{ref}=1900$.
Once evolved, the azimuthal structure is stable regarding an increase in electric potential.
Only for the largest \Ra$\;=4.7893\cdot 10^{4}$ for which we are able to count the azimuthal wavenumber, we observe an increase of $n$ if electric potential increases.
Here, electric gravity is high enough to modify the azimuthal structure.
If \Ra$\,$ increases one might also expect an increase of azimuthal wavenumber.
Indeed, this behaviour is observed.
Azimuthal wavenumber increases from $n=5$ to $n=7$.
But surprisingly, in the range in which one expects the wavenumber to be stable at
$n=5$ in a small range of \Ra$\;\approx1\cdot10^{4}$, it increases to wavenumber $n=6$.
With linear stability analysis~\citet{Meyer:2017a} calculated the critical azimuthal wavenumber (their fig. 2(c)) along the line of marginal stability (cf. also fig.~\ref{fig:regime}, thin solid line).
In an interval of $425\leq V_{0}/V_{ref}\leq3800$ the critical azimuthal wavenumber is $n_{c}=5$.
Aside from the fact that experimentally in a small range of \Ra, around $1\cdot10^{4}$ the azimuthal wavenumber surprisingly increases to $n=6$, we hereby validate the prediction of the critical wavenumber up to \Ra$\;=2.0526\cdot10^{4}$.
This is almost the Rayleigh number at which the line of marginal stability found by shadowgraph (fig.~\ref{fig:regime}, thick solid line) and PIV for $\Gamma=60$ (fig.~\ref{fig:regime}, dashed line) diverges strongly from theoretical prediction.

\subsection{Heat transfer}\label{sec:heat_transfer}
\begin{figure}
\centering
\includegraphics[width=0.49\textwidth]{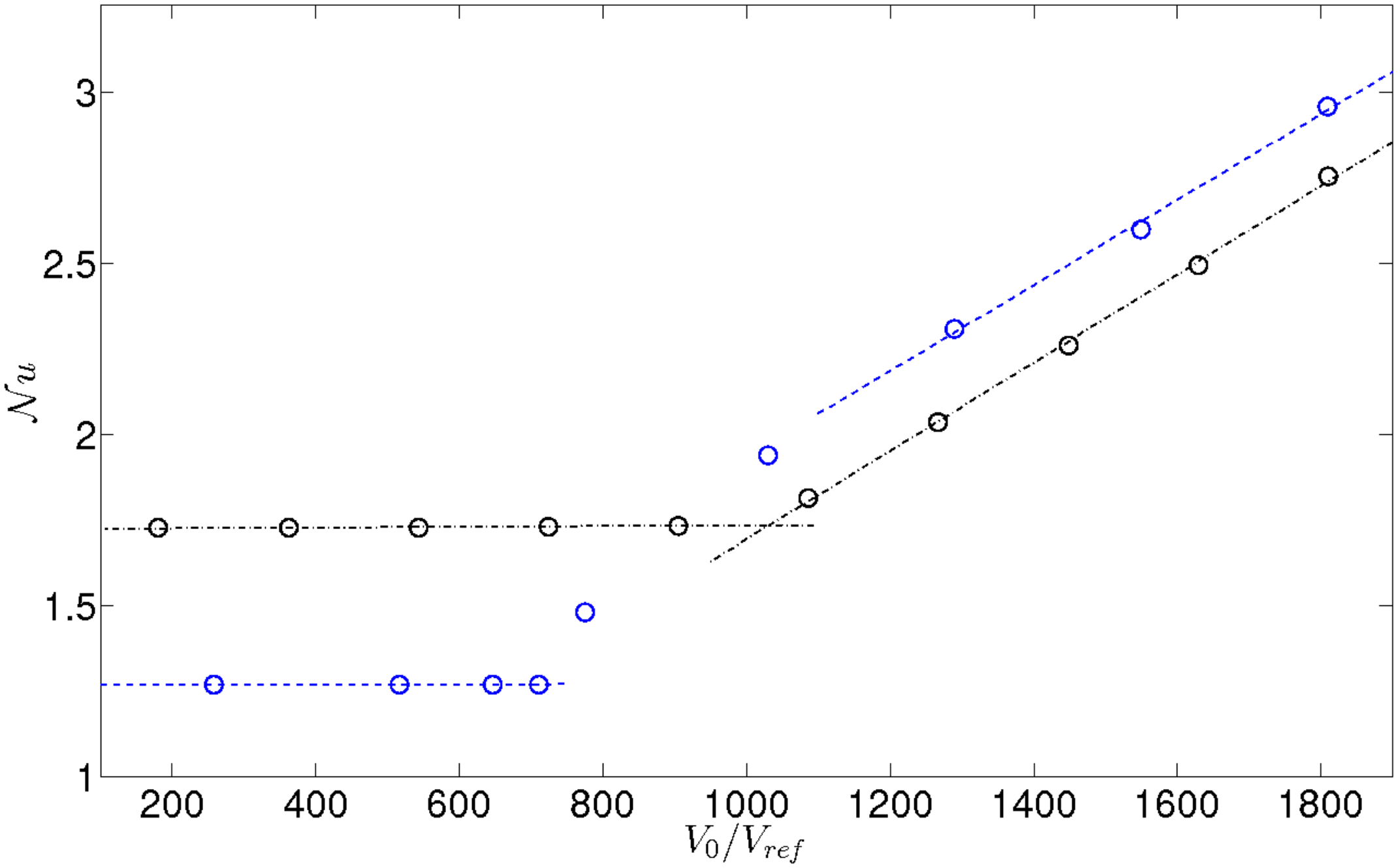}
\caption{Heat transfer in a vertical annulus of gap width $0.05\cdot d$, covering the area next to the inner wall. Between the inner and outer cylinder, a temperature difference of $\Delta T=7$ K (\Ra$\;=23946$) is applied. Numerical results are obtained for the experiment cell of aspect ratio $\Gamma=20$ (black circles) and aspect ratio $\Gamma=60$ (blue circles).}
\label{fig:Nu}
\end{figure}

The effectiveness of heat transfer is considered in terms of the Nusselt number.
We computed the Nusselt number in a vertical annulus of gap width $0.05\cdot d$, covering the area next to the inner wall (see also eq.~\eqref{eq:Nu}).
This choice turned out to yield results being numerically more stable than surface integrals over the inner wall.
We applied a temperature difference between the inner and outer cylinder of $\Delta T = 7$ K (\Ra$\;= 23946$) and increased the electric potential stepwise. 
The empirical linear relation between Nusselt number and dimensionless electric potential for aspect ratio $\Gamma=20$ (black line) and $\Gamma=60$ (blue line) is shown in figure~\ref{fig:Nu}.
In the case of the experiment cell with an aspect ratio of $\Gamma=20$ within the range of $1100 \leq V_{0}/V_{ref} \leq 1900$ ($5480\leq L \leq 15220$) where we found stationary axially aligned columnar structures we observed a linear dependency
\begin{eqnarray}
\mathcal{N}u & = & 1.29\cdot 10^{-3}V_{0}/V_{ref} + 0.403, \quad R^{2}=0.998,
\end{eqnarray}
where $R^{2}$ is the correlation coefficient.
In the case of the experiment cell with an aspect ratio of $\Gamma=60$ and taking only into account larger values of $V_{0}/V_{ref}\geq 1290$ ($L\geq 7765$), a linear dependency might also be appropriate
\begin{eqnarray}
\mathcal{N}u & = & 1.3\cdot 10^{-3}V_{0}/V_{ref} + 0.6858, \quad R^{2}=0.987,
\end{eqnarray}
although it does not predict the threshold of instability.
The fact that the aspect ratio plays an important role in the onset of DEP force-driven convection if the flow is in the convective regime is also visible in the Nusselt number.
It is valid that
\begin{eqnarray}
\left( \frac{V_{0,C}}{V_{ref}} \right)_{\Gamma = 20} > \left( \frac{V_{0,C}}{V_{ref}} \right)_{\Gamma = 60} > \left( \frac{V_{0,C}}{V_{ref}} \right)_{\Gamma = \infty}.
\end{eqnarray}

\section{Summary and Conclusions}\label{conclusions}
In the present paper, we studied dielectrophoretic force-driven convection in annular geometry under Earth's gravity.
Initially, to clarify the question if PIV is applicable to the flow, we investigated in a theoretical approach the influence of the dielectrophoretic force on particle movement. 
The theoretical study reveals that the movement of particles relative to the fluid is driven by two processes.
Firstly, due to the density difference and secondly due to the permittivity difference of the tracer particle and its surrounding fluid. 
The sedimentation velocity due to the density difference is in the order of $\mathcal{O}(10^{-3})$ mm/s.
The sedimentation velocity due to the permittivity difference is in the order of $\mathcal{O}(10^{-6})$ mm/s.
Both are negligible.
Therefore, PIV measurements with tracer particles of type Potters (HGS) are possible when they are mixed into a non-polar dielectric silicone oil and influenced by a high voltage a.c. electrical field.

With two experiment cells of different aspect ratio and a combination of two measurement techniques, shadowgraph (in case of apect ratio $\Gamma=20$) and PIV, we investigated the flow (i) without and (ii) with an electric potential applied.
In the first case, it is well known from experiments~\citep{Elder:1965a} that the flow in a vertical slot undergoes a transition from unicellular laminar (natural) free convection to a secondary flow (cat-eye pattern) for \Ra$_{C}\;= 3.6\cdot10^{5}$.
Substitution of the fluid and geometrical properties into the critical Rayleigh number, implies for our setup a radial temperature difference of about \mbox{$\Delta T\approx 88$ K}.
Additionally, there is a destabilising effect of radius ratio $\eta$ for high Prandtl numbers~\mbox{\cite[see][fig. 6]{Choi:1980}}.
\mbox{A maximal} applicable Rayleigh number of $2.053\cdot 10^{5}$ \mbox{($\Delta T = 60$ K)} for both of the experiment cells and the weakly destabilising effect of aspect ratio for high \Pra$\,$ seems to small to find cat-eye pattern.
\citet{Meyer:2017a} (fig.~\ref{fig:regime}, thin solid line) performed linear stability analysis for a cylindrical annulus of infinite length and considered the same physical and geometrical properties as we use in this study.
In absence of an electric potential, they found for \Ra$\;=5.009\cdot10^{4}$ \mbox{($\Delta T = 14.64$ K)} critical modes in the form of oscillatory axisymmetric vortices (vertical wavenumber $k_{C}\approx2.5$) of the thermal instability~\cite[see also][]{Bahloul:2000}.
They considered the temperature profile and the vertical velocity profile, which corresponds to the conductive regime~\citep{Choi:1980,Lopez:2015,Meyer:2017a}.
Near the midheight the flow is parallel to the vertical sidewalls.
Heat is transferred across the gap by conduction alone.
The flow is said to be in the conductive regime.
The parallel flow extends towards the end-plates the smaller the Rayleigh number or the higher the aspect ratio~\citep{Choi:1980}.
\cite{Vahl_Thomas:1969} showed, if the Prandtl number is sufficiently large and the aspect ratio low the vertical flow begins to take the form of boundary layers near the side walls.
'The flow enters the convective regime in which, in addition to the boundary layers, a stable vertical temperature gradient develops in the core of the flow'~\citep{Choi:1980}.
Referring to~\citet{GillDavey:1969}, the conductive regime is realised if \Ra$_{C}<300\Gamma$.
\citet{Lopez:2015} recently extended the criterion to \Ra$_{C}(\eta,\Gamma)<a(\eta)\Gamma$, with $a(\eta=0.5)=329.375$.
The criterion determines whether the model used for linear stability analysis considering an infinite extend in the axial direction is a good approximation of the laboratory flow.
For our experimental setup with aspect ratio $\Gamma=20$ or 60 the criterion gives \Ra$_{C}<6587$ or $1.963\cdot 10^{4}$, (cf. also fig.~\ref{fig:regime} horizontal dot-dashed and dashed line).
~\citet{Elder:1965b} found the empirical relation \Ra$_{C}=8\cdot10^{8}\mathcal{P}r^{1/2}/\Gamma$.
Rearranging the relation and substitution of \Ra$_{C}$ found with the criterion from~\citet{Lopez:2015} yields to $\Gamma_{C} \geq 66.5$.
From all the considerations we draw the following conclusions.
The flow is in the convective regime, which modifies the base temperature as well as the base vertical velocity profile and hence stabilises the flow.
The aspect ratio is too small to find the secondary flow superposed on the unicellular base flow.
Indeed, we did not observe a secondary flow.
Hence, initial condition for experiments with an electric potential applied is the unicellular laminar (natural) free convection.

We also investigated experiments with an electric potential applied in a wide range of tuples of ($V_{0}/V_{ref}$,\Ra) and drew a regime diagram regarding their stability.
We identified lines of marginal stability regarding aspect ratio and measurement technique and compared them with the marginal curve found by linear theory~\citep{Meyer:2017a}.
In a range of \Ra$\;<6800$ we found the transition from stable to unstable flow independent of aspect ratio.
Hence we validate theoretical prediction of the linear stability analysis.
 
We showed that the larger the aspect ratio is, the closer the transition is to the prediction of linear stability analysis.  
It substantiates the transition to be linear in an interval ($V_{0}/V_{ref}$,\Ra) where the Boussinesq approximation is valid and may also give a threshold, i.e. for \Ra$\;\approx 2.5\cdot 10^{4}\equiv\Delta T \approx 7.3$ K, where the break-up occurs and adiabatic end-plate effects and/or non-linear processes have to be considered.

Since stable flow is referred to unicellular laminar (natural) free convection the criterion found by~\citet{GillDavey:1969} and recently extended by~\citet{Lopez:2015} seems also valid.
Interestingly, the break-up of the line of marginal stability occurs in close proximity to the criterion.
The break-up in case of $\Gamma=20$ is at \Ra$_{C}\approx 6800$ and in case of $\Gamma=60$ at \Ra$_{C}\approx 2.4\cdot 10^{4}$.
There the conductive regime enters the convective regime.
Once the convective regime sets in, the transition from unicellular laminar free convection to thermal electro-hydrodynamic convection is independent of electric potential.

After the first transition, the flow converges during a transient phase upon switching on electric potential always to a stationary 3D structure/flow. 
We identified by the combination of shadowgraph technique, PIV and numerical simulation the structure as axially aligned stationary columns.
They consist of equidistant, azimuthally alternating radially in/outward directed convective plumes trapped between upper and lower boundary layer circulations.

The transient phase develops always in the same manner.
In the radial-vertical plane, the flow destabilises first in the lower and midheight part of the gap.
The shear-layer interface separating downward flow (colder fluid) from upward (warmer fluid) flow becomes wavy.
The perturbation in the lower part wins the competition and the structure grows upward until it reaches a new equilibrium state.

The study concerning quasi-stationary increase/decrease of electric potential reveals the azimuthal position of radially inward/outward directed plumes depends strongly on electric potential. 
An increase of electric potential shifts the instability counterclockwise until electric potential and hence electric gravity is high enough and the pattern becomes arrested in its azimuthal position.
The shift is approximately $\pi/2$ and decreases rapidly when \Ra$\,$ increases.
Contrary, if electric potential quasi-stationary decreases the pattern first shifts its azimuthal position clockwise ending in an arrested state before the transition to stable unicellular laminar free convection.
The transition itself is not subjected to hysteresis.

From shadowgraph measurements, we finally drew a regime diagram in terms of azimuthal wavenumber $n$ spanned by dimensionless electric potential and Rayleigh number.
In the range of ($V_{0}/V_{ref}$,\Ra) after the first transition, where the flow is columnar and stationary and where we are able to count the azimuthal wavenumber, $n$ once evolved is insensitive regarding an increase of electric potential.
Only at the largest \mbox{\Ra$\;=4.7893\cdot10^{4}$} we observe an increase of $n$ when electric potential increases.
In a diagram spanned by dimensionless electric potential and Rayleigh number each tuple ($V_{0}/V_{ref}$,\Ra) is equidistantly spaced.
Taking into account that the electric Rayleigh number depends on the temperature difference as well as on the electric potential (cf. eq.~\eqref{eq:eg}), then there appears non-uniform spacing between tuples.
The artificial electric gravity and hence $L$ increase with \Ra.
For \Ra$\;$ sufficiently small, this suggests that the azimuthal wavenumber is ’insensitive’ to an increase of the electric potential.
Contrary, the azimuthal wavenumber continuously increase with \Ra.
But surprisingly, in a range expecting a stable wavenumber $n=5$ it increases to $n=6$ in a small band of \Ra$\;\approx1\cdot10^{4}$.  
Its origin is speculative and remains an open question.
\citet[][fig 2(c)]{Meyer:2017a} found in an interval $425\leq V_{0}/V_{ref}\leq 3800$ the critical azimuthal wavenumber $n_{c} = 5$ along the line of marginal stability.
Aside from the fact that experimentally in a small range of \Ra, the azimuthal wavenumber surprisingly jumps, we hereby validate not only the prediction of the critical wavenumber up to \Ra$\;=2.0526\cdot 10^{4}$ but also the prediction of the stationary axially aligned columnar structure (\cite{Meyer:2017a}, fig. 2(b,e) critical vertical wavenumber $k_{c}=0$ and critical frequency $\omega_{c}=0$). 

The analysis of heat transfer in terms of the Nusselt number reveal, as long as the flow is unicellular, an increase of electric potential has a weak effect on heat transfer. 
Obviously, the closer the aspect ratio to the infinite vertical length approximation, the closer the Nusselt number to the pure conduction case.
Once the flow enters the regime of stationary columnar vortices ($V_{0}/V_{ref} \geq 1100$, $L \geq 5480$) a linear dependency for \Nu$\;$ on electric potential was found for aspect ratio $\Gamma=20$.
In case of aspect ratio $\Gamma=60$ the flow enters the first regime when $V_{0}/V_{ref} \geq 780$, $L \geq 2796$.
The Nusselt number grows in close proximity to the critical value. 
Taking only into account larger values of $V_{0}/V_{ref}$ a linear dependency might be appropriate.
Both curves show a significant enhancement of the heat transfer through the inner cylindrical annulus.
The critical values can also be used for comparison with their experimental counterparts, $V_{0}/V_{ref}\approx 1000$ ($\Gamma=20$, shadowgraph), $V_{0}/V_{ref}\approx 1150$ ($\Gamma=20$, PIV) and $V_{0}/V_{ref}\approx 780$ \mbox{($\Gamma=60$, PIV)}.

\section{Acknowledgements}\label{acknowledgements}
This work was supported by the German Research Foundation (DFG) grant "Thermo-elektro-hydrodynamisch TEHD getriebene W\"{a}rmetransporterh\"{o}hung im vertikalen Zylinderspalt - Experimente und numerische Simulation im Kontext von Messunsicherheiten und optimaler Versuchsplanung (EG 100/20-1)``.
M. Jongmanns, A. Meyer and M. Meier acknowledge the support of the BMWi via German Aerospace Center DLR under grant no. 50WM1644.
P. Gerstner, M. Baumann and V. Heuveline acknowledge support by the state of Baden-W\"{u}rttemberg through bwHPC and the DFG through grant INST 35/1134-1 FUGG.
We further thank Markus Helbig and Vilko Ostmann for technical assistance.
We also thank the anonymous reviewers for their critical remarks that helped to improve the quality of the final version. \\



\bibliographystyle{elsarticle-harv} 
\bibliography{TEHD}





\end{document}